\documentclass[draftcls,onecolumn]{IEEEtran}
\ifCLASSINFOpdf
   \usepackage[pdftex]{graphicx}
\else
\fi

\usepackage{graphicx}
\usepackage{epsf}
\usepackage{epsfig}
\usepackage{cite}
\usepackage{color}
\usepackage{amsmath,amssymb,amsfonts,mathtools}

\newtheorem{theorem}{Theorem}
\newtheorem{proposition}{Proposition}

\begin{document}
%
\title{A Finite-Blocklength Perspective \\ on Gaussian Multi-Access Channels}
%
%
%

\author{Ebrahim~MolavianJazi,~\IEEEmembership{Student~Member,~IEEE,}
        and~J.~Nicholas~Laneman,~\IEEEmembership{Senior~Member,~IEEE}
\thanks{Ebrahim MolavianJazi and J. Nicholas Laneman are with the Department
of Electrical Engineering, University of Notre Dame, Notre Dame,
IN, 46556 USA e-mails: \{emolavia,jnl\}@nd.edu.}
}

\maketitle


\begin{abstract}
Motivated by the growing application of wireless multi-access networks with stringent delay constraints, we investigate the Gaussian multiple access channel (MAC) in the finite blocklength regime. Building upon information spectrum concepts, we develop several non-asymptotic inner bounds on channel coding rates over the Gaussian MAC with a given finite blocklength, positive average error probability, and maximal power constraints.  Employing Central Limit Theorem (CLT) approximations, we also obtain achievable second-order coding rates for the Gaussian MAC based on an explicit expression for its dispersion matrix. We observe that, unlike the pentagon shape of the asymptotic capacity region, the second-order region has a curved shape with no sharp corners.  

A main emphasis of the paper is to provide a new perspective on the procedure of handling input cost constraints for tight achievability proofs. Contrary to the complicated achievability techniques in the literature, we show that with a proper choice of input distribution, tight bounds can be achieved via the standard random coding argument and a modified typicality decoding. In particular, we prove that codebooks generated randomly according to \emph{independent uniform distributions on the respective ``power shells''} perform far better than both independent and identically distributed (i.i.d.) Gaussian inputs and TDMA with power control. Interestingly, analogous to an error exponent result of Gallager, the resulting achievable region lies roughly halfway between that of the i.i.d. Gaussian inputs and that of a hypothetical ``sum-power shell'' input. However, dealing with such a non-i.i.d. input requires additional analysis such as a new \emph{change of measure} technique and application of a Berry-Esseen \emph{CLT for functions} of random variables.

\end{abstract}



%
\IEEEpeerreviewmaketitle

\section{Introduction}
\label{Sec-Intro}

%
%
%
%

\IEEEPARstart{W}{ireless} multi-access networks are increasingly emerging as an integral part of many communication and control systems with a central data processing or decision making unit, such as the uplink of wireless cellular communications, sensor networks, and machine-to-machine (M2M) communication systems. Such multi-access communication networks usually have low latency constraints for their information, due to their nature or application. These delay requirements, along with the desire for low-complexity system designs, call for schemes and protocols that employ finite blocklengths, even on the order of several hundred symbols, and achieve high levels of reliability at the same time.

A mathematical analysis and design of multi-access networks with such stringent latency requirements, however, cannot rely on conventional information theoretic results, which assume asymptotically large blocklengths and vanishingly small error probability. It is therefore critical to develop rigorous non-asymptotic results that are tight for finite blocklengths. Although this has been an strong trend in the early ages of information theory~\cite{Shannon,Strassen}, there has been renewed interest in this direction since the landmark works of~\cite{PPV,Hayashi}. The main theme of these works is treating mutual information as a random variable (RV)~\cite{Fano, JNL}, which has a stochastic behavior based on the transmitted input and the channel noise and interference. This idea is mainly developed in the information spectrum approach of Verd\'u and Han~\cite{Verdu-Han,Han}, which suggests that the cumulative distribution function (CDF) of this RV characterizes performance in terms of the probability that the channel cannot support the communication rate and causes an ``outage'' for the actual codeword to be correctly detected at the receiver. The highest coding rates arise if the error probability is dominated by the outage probability, and the probability of ``confusion'', i.e., the observation is wrongly decoded to any incorrect codeword, decays to zero. 

Although tight non-asymptotic bounds obtained by the information spectrum approach can help with precise analysis and design of communication systems, their numerical computation are usually cumbersome. It is therefore of high practical interest to come up with accurate approximations of the coding rates that are still valid for moderately short blocklengths. Capacity (region), as a first-order statistic of the channel, is already a first-order approximation of the coding rates, but it is only useful for very long blocklengths. Error exponent~\cite{Gallager,Csiszar} is one conventional tool for this purpose, which applies Large Deviation Theory (LDT) to the mutual information RV and studies the exponential decay in error probability of a fixed-rate coding scheme as the blocklength grows larger. Although error exponent analysis can provide a rough estimate for finite-blocklength analysis, a method for finding sharper approximations employs the Central Limit Theorem (CLT) to the mutual information RV, specifically for rates close to capacity. This way, one can investigate the increase in coding rate of a scheme with fixed error probability as the blocklength grows larger and obtain second (and higher) order approximations. In particular, it has been demonstrated~\cite{Strassen,PPV,Hayashi} that second-order approximations involving the fundamental quantity of \emph{channel dispersion}, as a second order statistic of the channel, provide good estimates of the channel coding rates for moderate to short blocklengths.

In this paper, we show how similar ideas can be extended to a multi-user setting in which multiple users are communicating several independent messages to a single receiver over a Gaussian multiple access channel (MAC). In particular, we present several non-asymptotic achievability bounds on the channel coding rates of a Gaussian MAC as a function of the finite blocklength, the fixed average error probability, and the users' power constraints. Our bounds suggest that the \emph{joint outage} event, in which either of the users' mutual information is not strong enough to support its target rate, is the fundamental quantity that governs the performance over the Gaussian MAC. Since this joint outage event is in general complex, we also give a slightly looser, but simpler to analyze non-asymptotic achievable region based on an \emph{outage-splitting} idea~\cite{ML-ISIT12}, in which the joint outage event is split into individual outage events via the union bound. 
Applying the CLT to our finite-blocklength results, we also obtain corresponding achievable second-order coding rate regions for the Gaussian MAC. In particular, we give explicit expressions for the achievable dispersion matrices of the Gaussian MAC in terms of the users' power constraints.

A critical ingredient of our analysis is the choice of input distribution for improving the second-order performance. 
In particular, consider the 2-user Gaussian MAC with maximal power constraints $P_1$ and $P_2$. Inspired by Shannon~\cite{Shannon}, rather than random coding using the common choice of independent and identically distributed~(i.i.d) Gaussian input distributions $X^n_1\overset{\text{i.i.d.}}{\sim}\mathcal{N}(0,P_1),X^n_2\overset{\text{i.i.d.}}{\sim}\mathcal{N}(0,P_2)$ ~\cite{Rice,Cover}, which achieve the  capacity region and are therefore optimal to first order, or their truncated versions lying in thin shells $nP_1\!-\delta \leq\!||x_1^n||^2\!\leq nP_1, \; nP_2-\delta \leq\!||x_2^n||^2\leq nP_2$ for an arbitrarily small $\delta>0$, which are used by Gallager for the error exponent analysis~\cite{Gallager,Gallager-MAC},  we focus on inputs having \emph{independent uniform distributions on the respective power shells}, namely, the $n$-dimensional spheres~$||x_1^n||^2=nP_1$ and $||x_2^n||^2=nP_2$.

Consider a symmetric Gaussian MAC with blocklength~$n=500$, average error probability $\epsilon=10^{-3}$, and powers $P_1\!=\!P_2=0$~dB. Figure 1 compares\footnote{This is a corrected and updated version of a similar plot which was presented in the conference version of this work~\cite{ML-Allerton12}.} the approximate achievable rate regions for all of the aforementioned input distributions: independent power-shell inputs with both joint-outage and outage-splitting versions; independent i.i.d. Gaussian inputs; independent truncated Gaussian inputs; and also the rate region achievable via time division multiple access (TDMA) with power control; along with the asymptotic Cover-Wyner capacity region~\cite{Cover-GMAC,Wyner}. We also depict a \textit{hypothetical} rate region which would be achievable if the sum of independent power shell inputs fell on the sum-power shell.\footnote{We conjecture this to be an outer bound for the Gaussian MAC.} To show the tightness of the achievable rate regions, we also depict two straightforward second-order single-user (SU) outer bounds and a conjectured second-order sum-rate outer bound. The details of all of these regions are given in Section IV. We note that all of the approximation results are computed only up to second-order.

\begin{figure}
\label{fig-low}
\begin{center}
\includegraphics[scale=0.65]{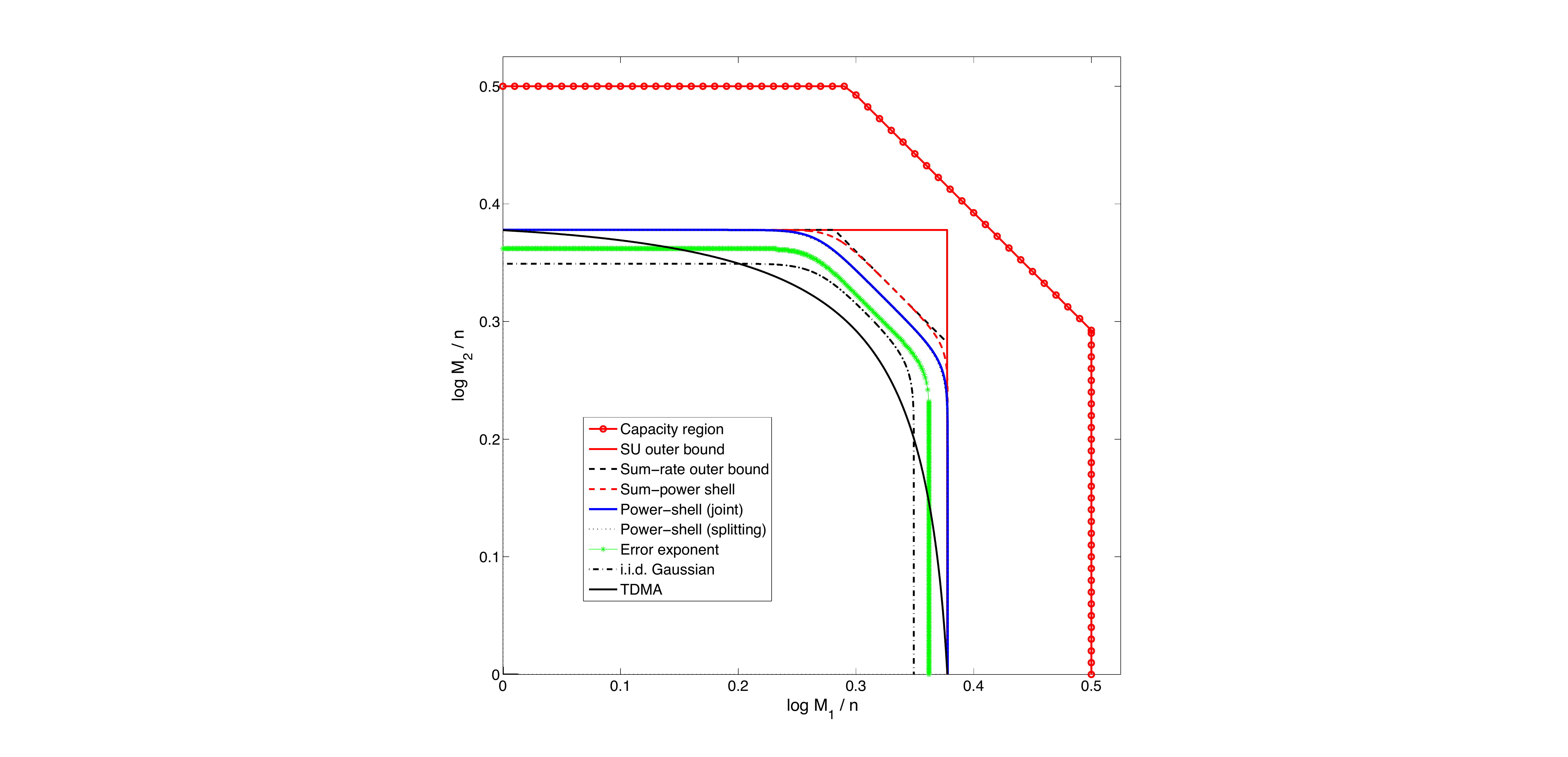}
\end{center}
\caption{Symmetric Gaussian MAC with blocklength~$n=500$, average error probability $\epsilon=10^{-3}$, and powers $P_1\!=\!P_2=0$~dB.}
\end{figure}


%

Unlike the pentagon shape of the capacity region of the Gaussian MAC in the infinite blocklength regime, we observe that its second-order approximation has a curved shape with no sharp corners. Moreover, the region resulting from independent power shell inputs lies roughly halfway between that of the i.i.d. Gaussian inputs and that which would be achievable by the hypothetical ``sum-power shell'' input. This phenomenon is similar to one observed by Gallager in his analysis of error exponents for the Gaussian MAC~\cite{Gallager-MAC}. It is also interesting that, Gaussian (and truncated Gaussian) random codebooks, although optimal for achieving capacity, are not second-order optimal, and their finite blocklength achievable rate region falls well inside that of power shell inputs. Another interesting observation is that, contrary to the infinite blocklength case, the TDMA strategy with power control is not even sum-rate optimal. Last but not least, the outage-splitting region of the power-shell input closely resembles that of the joint-outage version, and therefore its simplicity does not sacrifice much with respect to accuracy.

Of course, the improved performance of the independent power shell inputs comes at the price of additional complexity in the analysis. First, although the variance of the sum is the sum of variances for two independent Gaussians, the sum of two independent power shell inputs does not lie on the power shell corresponding to the sum of the powers, i.e., $||x_1^n||^2+||x_2^n||^2 \neq ||x_1^n+x_2^n||^2$. Second, classical CLT and LDT analysis do not apply directly for such non-i.i.d. inputs. To overcome these difficulties, we develop new techniques: since power shell inputs can be constructed by normalizing i.i.d. Gaussian RVs, we rely on a \textit{CLT for functions} to develop the outage probability approximation; additionally, we introduce a \textit{change of measure} technique for the confusion probability, so that classical LDT can be applicable to prove its decay to zero.

Another main emphasis of our work has been to utilize standard and transparent methods to highlight the proof steps for finite blocklength analysis, especially when input cost constraints are involved. We are specifically focused on the method of encoding and decoding. Although \emph{random coding and typicality decoding} have proven to be powerful tools in information theory and the standard method for proving most source and channel coding theorems~\cite{ElGamal-Kim}, all non-asymptotic achievability bounds for the Gaussian channel either use random coding but with maximum likelihood (ML) decoding~\cite{Shannon,Gallager}, or employ typicality decoding but with non-random sequential encoding~\cite{PPV,Thomasian}. In addition, for the tightest bounds, handling the cost constraint is either done through relatively sophisticated geometric arguments~\cite{Shannon} or via a relatively complicated introduction and analysis of composite hypothesis testing~\cite{PPV}. In this paper, we start by proving tight finite-blocklength achievability results for point-to-point (P2P) Gaussian channels, which are at least second-order optimal, using the standard arguments of random coding and typicality decoding, with some slight modifications. As we will see, this approach appears easier to generalize than those in~\cite{Shannon,PPV} to multi-user settings, specifically the Gaussian MAC, for which we obtain rather tight achievable second-order approximations using the random coding and modified typicality decoding method.

The rest of this paper is organized as follows. In Section~\ref{Sec-BG}, we review the tightest achievability methods in the finite blocklength regime. Then, in Section~\ref{sec-P2P}, to highlight the key elements of our proof techniques, we revisit the problem of the P2P Gaussian channel, develop new non-asymptotic achievability bounds, and re-derive the second-order approximation of~\cite{PPV,Hayashi}. In Section~\ref{sec-MAC}, we turn to to our problem of interest, prove finite-blocklength inner bounds for the Gaussian MAC, and then apply them to establish achievable second-order coding rates. We conclude the paper in Section~\ref{sec-con} and relegate some of the technical proofs to the Appendices.



%
%

\section{Background on Tight Achievability Methods}
\label{Sec-BG}

To highlight the conciseness and simplicity of our approach in proving accurate non-asymptotic and approximate achievability results for cost-constrained channel models, specifically the P2P Gaussian channel and the Gaussian MAC, in this section we review the sharpest and most well-known achievability methods in the literature.


We first review the details of random coding and typicality decoding for proving the achievability side of the coding theorems, because of their simplicity and also because we slightly modify these methods to prove sharp achievability bounds for Gaussian (and other cost-constrained) channels. To emphasize the transparency and conciseness of this approach, we will then review the details of two of the sharpest bounds for the Gaussian channel, namely Polyanskiy et al.'s $\kappa\beta$ method based on composite hypothesis testing~\cite{PPV} and Shannon's geometric method~\cite{Shannon}, and point out the complexities of these methods and the difficulties in generalizing them to multi-user settings. We explain these methods to some level of details to highlight some of their key tools and concepts that we leverage in our later analysis. Note that, in this section, we are concerned with non-asymptotic achievability bounds that are valid for any finite blocklength without requiring convergence conditions.

\subsection{Random Coding and Typicality Decoding}
\label{sub-RC-TD}

The basic idea in an argument based upon random coding and typicality decoding can be reviewed most clearly for a P2P channel~$P_{Y^n|X^n}(y^n|x^n)$. The channel encoder randomly generates $M$~codewords $\{x^n(j)\}_{j=1}^M$ of the codebook independently according to some given $n$-letter distribution $P_{X^n}(x^n)$, where $n$ is the designated blocklength. Observing the output $y^n$, the decoder then chooses the first codeword $x^n(\hat{m})$ of the codebook which looks ``typical'' with $y^n$ in a one-sided sense\footnote{    The use of ``typicality'' nomenclature for this threshold decoding is inspired by the \textit{two-sided} threshold decoding in conventional typicality definition, e.g. by Cover and Thomas~\cite[Section 8.6]{Cover}: $(x^n,y^n)$ are jointly typical if 
\[
\left|\frac{1}{n}\log P_{X^n}(x^n)-\mathbb{H}(X)\right|<\epsilon, \quad \left|\frac{1}{n}\log P_{Y^n}(y^n)-\mathbb{H}(Y)\right|<\epsilon, \quad \left|\frac{1}{n}\log P_{X^nY^n}(x^n,y^n)-\mathbb{H}(X,Y)\right|<\epsilon,
\]
and Han~\cite[Section 3.1]{Han}: $(x^n,y^n)$ are jointly typical if 
\[
\left|\frac{1}{n}\log \frac{P_{Y^n|X^n}(y^n|x^n)}{P_{Y^n}(y^n)}-\mathbb{I}(X;Y)\right|<\gamma,
\]
where $\mathbb{H}(X)$ and $\mathbb{I}(X,Y)$ denote the average entropy and the average mutual information, respectively. Note that the latter condition of Han is implied by the former set of conditions of Cover and Thomas.
}
\begin{equation}
i(x^n(\hat{m});y^n)>\log \gamma(x^n(\hat{m})),
\end{equation}
where $\gamma(x^n)$ is a (possibly) codeword-dependent threshold and $i(x^n;y^n)$ is the corresponding realization of the \emph{mutual information} RV 
\begin{equation}
i(X^n;Y^n):=\log \frac{P_{Y^n|X^n}(Y^n|X^n)}{P_{Y^n}(X^n)}.
\label{inf-RV}
\end{equation}
Here, the \emph{reference} distribution $P_{Y^n}$ is the marginal output distribution induced by the input distribution~$P_{X^n}$, i.e., 
\begin{equation}
P_{Y^n}(y^n)=\sum_{x^n\in\mathcal{X}^n} P_{X^n}(x^n) P_{Y^n|X^n}(y^n|x^n).
\end{equation}

Using one realization of such a code $\{x^n(j)\}_{j=1}^M$, the average error probability can be bounded as\footnote{Throughout this paper, we use a non-standard notation of the form~$P_XP_YP_{Z|X}[f(X,Y,Z)\in\mathcal{A}]$ to explicitly indicate that $(X,Y,Z)$ follow the joint distribution~$P_X\times P_Y\times P_{Z|X}$ in determining the probability $\Pr[f(X,Y,Z)\in\mathcal{A}]$.}
\begin{align}
\epsilon &\leq \frac{1}{M}\sum_{k=1}^M P_{Y^n|X^n=x^n(k)}[i(x^n(k);Y^n)\leq \log \gamma(x^n(k))] \notag \\
& \quad + \frac{1}{M}\sum_{k=1}^M P_{Y^n|X^n=x^n(k)}\left[\bigcup_{j=1}^{k-1}i(x^n(j);Y^n)>\log \gamma(x^n(j))\right], \label{typ-1}
\end{align}
that is, the sum of an \emph{outage probability}, that the correct codeword does not look typical, and a \emph{confusion probability}, that a preceding codeword incorrectly looks typical. 

The error probability averaged over all possible realizations of the codebook can then be bounded as
\allowdisplaybreaks{
\begin{align}
\epsilon &\leq \prod_{l=1}^M\left[ \sum_{x^n(l)} P_{X^n}(x^n(l))\right] \frac{1}{M}\sum_{k=1}^M  P_{Y^n|X^n=x^n(k)}[i(x^n(k);Y^n)\leq \log \gamma(x^n(k))] \notag \\
& \;\;\; +\prod_{l=1}^M\left[ \sum_{x^n(l)} P_{X^n}(x^n(l))\right] \frac{1}{M}\sum_{k=1}^M P_{Y^n|X^n=x^n(k)}\left[\bigcup_{j=1}^{k-1}i(x^n(j);Y^n)>\log \gamma(x^n(j))\right] \label{ave} \\
&\leq  \frac{1}{M}\sum_{k=1}^M  \sum_{x^n(k)} P_{X^n}(x^n(k)) P_{Y^n|X^n=x^n(k)}[i(x^n(k);Y^n)\leq \log \gamma(x^n(k))] \prod_{l\neq k}\left[ \sum_{x^n(l)} P_{X^n}(x^n(l))\right] \notag \\
& \;\;\; +  \frac{1}{M}\sum_{k=1}^M \sum_{j=1}^{k-1} \sum_{x^n(j)} \sum_{x^n(k)} P_{X^n}(x^n(j)) P_{X^n}(x^n(k)) P_{Y^n|X^n=x^n(k)}\left[i(x^n(j);Y^n)>\log \gamma(x^n(j))\right]  \prod_{l\neq j,k}\left[ \sum_{x^n(l)} P_{X^n}(x^n(l))\right]  \label{uni} \\
&=  \frac{1}{M}\sum_{k=1}^M  P_{X^n}P_{Y^n|X^n}[i(X^n;Y^n)\leq \log \gamma(X^n)] +  \frac{1}{M}\sum_{k=1}^M (k-1) P_{X^n}P_{Y^n}[i(X^n;Y^n)>\log \gamma(X^n)] \\
&\leq P_{X^n}P_{Y^n|X^n}[i(X^n;Y^n)\leq \log \gamma(X^n)] + \frac{M-1}{2} P_{X^n}P_{Y^n}[i(X^n;Y^n)>\log \gamma(X^n)], \label{typ-2}
\end{align}}
where~\eqref{ave} follows from averaging over the random codebook, and~\eqref{uni} follows from the union bound.

The final result is that there exists a deterministic codebook consisting of $M$ codewords whose average error probability~$\epsilon$ satisfies~\eqref{typ-2}. It is worth mentioning that, in the standard asymptotic analysis of memoryless channels $P_{Y^n|X^n}(y^n|x^n)=\prod_{t=1}^n P_{Y|X}(y_t|x_t)$, the input distribution is selected i.i.d. $P_{X^n}(x^n)=\prod_{t=1}^n P_X(x_t)$, and the threshold is selected as a function of the \emph{average mutual information}, $\log\gamma(x^n)=\log \gamma_n=n\mathbb{I}(X;Y)-o(n)=n\mathbb{E}_{P_XP_{Y|X}}[i(X;Y)]-o(n)$. This leads to the proof of achievability for rates $\frac{\log M}{n}<\mathbb{I}(X;Y)$. In this paper, however, we preserve the general $n$-letter form of the input distribution, since we will use non-i.i.d. input distributions to deal with input cost constraints.

The result~\eqref{typ-2} can be readily extended to input cost constrained settings~\cite{Thomasian} requiring~$X^n\in\mathcal{F}_n$, where $\mathcal{F}_n\subset \mathcal{X}^n$ is the set of feasible input sequences: upon selecting the decoding threshold
\begin{align}
\gamma(x^n)=
\begin{cases}
\gamma_n & x^n\in\mathcal{F}_n\; , \\
\infty & x^n\notin\mathcal{F}_n\; ,
\end{cases} \label{inf-th}
\end{align}
where $\gamma_n$ is a prescribed threshold, we obtain
\begin{align}
\epsilon &\leq P_{X^n}P_{Y^n|X^n}\left[{i}(X^n;Y^n)\leq \log \gamma(X^n) \bigcap X^n\notin\mathcal{F}_n\right]  + P_{X^n}P_{Y^n|X^n}\left[{i}(X^n;Y^n)\leq \log \gamma(X^n) \bigcap X^n\in\mathcal{F}_n\right] \notag \\
& \quad + \frac{M-1}{2} P_{X^n}P_{Y^n}\left[{i}(X^n;Y^n)>\log \gamma(X^n)\bigcap X^n\notin\mathcal{F}_n\right]  \notag \\
& \quad + \frac{M-1}{2} P_{X^n}P_{Y^n}\left[{i}(X^n;Y^n)>\log \gamma(X^n)\bigcap X^n\in\mathcal{F}_n\right]  \label{DT-cost-1} \\
&\leq P_{X^n}[X^n\notin\mathcal{F}_n] + P_{X^n}P_{Y^n|X^n}[i(X^n;Y^n)\leq \log \gamma_n] + \frac{M-1}{2} P_{X^n}P_{Y^n}[i(X^n;Y^n)>\log \gamma_n]. \label{DT-cost-PPV}
\end{align}
Upon remapping all non-feasible codewords to an arbitrary sequence~$x^n(0)\in\mathcal{F}_n$ and without touching the decoding regions, we conclude that there exists a deterministic codebook with $M$ codewords all belonging to the feasible set~$\mathcal{F}_n$ and whose average error probability~$\epsilon$ satisfies~\eqref{DT-cost-PPV}, cf.~\cite[p. 2314]{PPV}.

Considering an i.i.d. Gaussian input $P_{X^n}\sim\mathcal{N}(\mathbf{0},(P-\delta)I_n)$, with $\delta$ being any arbitrarily small positive constant\footnote{The power margin $\delta$ can be vanishing with $n$ provided that it decays strictly slower than $O\left(\frac{1}{\sqrt{n}}\right)$ so that the cost-violation probability does not dominate.}, and applying the conventional CLT to~\eqref{DT-cost-PPV} results in the approximate achievability bound~\cite{Rice}:
\begin{align}
\frac{\log M}{n}\leq C(P-\delta)-\frac{\log e}{\sqrt{n}}\sqrt{\frac{P-\delta}{1+P-\delta}}Q^{-1}(\epsilon)+O\left(\frac{1}{n}\right), \label{Gaussian-input}
\end{align}
where, as usual, $Q^{-1}(\cdot)$ is the functional inverse of the complementary cumulative distribution function (CDF) of a standard Gaussian distribution $Q(x)=\frac{1}{\sqrt{2\pi}}\int_x^\infty e^{-t^2/2}dt$, and
\begin{equation}
C(P)=\frac{1}{2}\log(1+P).
\label{capacity}
\end{equation}
As will be seen shortly, this second-order performance is not optimal. Therefore, the i.i.d. Gaussian input distribution achieves capacity but is not second-order optimal, since a considerable portion of Gaussian codewords do not utilize the maximum available power budget~$P$, which degrades the performance. In Shannon's words~\cite{Shannon}, ``it is evidently necessary to avoid having too many of codepoints interior to the $\sqrt{nP}$ sphere.'' It will be shown that more refined input distributions, that force all codewords to use the maximum power~$P$, are required for this purpose.

\subsection{Polyanskiy et al.'s $\kappa\beta$ Bound}
\label{sub-kap-beta}

A tighter achievability result for the P2P Gaussian channel is provided in the recent $\kappa\beta$~bound of Polyanskiy et al.~\cite{PPV}. Using a slightly different language from that in~\cite{PPV}, this bound fixes an arbitrary \emph{output} distribution $Q_{Y^n}$, similar to~\cite{Fano,Wolfowitz}, and employs this as the reference distribution for the definition of a \emph{modified} mutual information random variable:
\begin{equation}
\tilde{i}(X^n;Y^n):=\log \frac{P_{Y^n|X^n}(Y^n|X^n)}{Q_{Y^n}(Y^n)}.
\end{equation}
Building upon the maximal coding idea~\cite{Feinstein,Ash}, deterministic sequences are arbitrarily chosen as codewords one by one, and the sequential codeword generation process stops after selecting $M$ codewords $\{x^n(j)\}_{j=1}^M$ if the error probability for any choice of the $(M+1)$-th sequence exceeds the target maximal error probability~$\epsilon$, i.e.,
\begin{align}
\epsilon &< P_{Y^n|X^n=x^n}[\,\tilde{i}(x^n;Y^n)\leq\log \gamma_n] + P_{Y^n|X^n=x^n}\left[\bigcup_{j=1}^{M} \tilde{i}(x^n(j);Y^n)>\log \gamma_n\right] \label{kappa-1}
\end{align}
for all sequences $x^n\in\mathcal{F}_n$, where $\mathcal{F}_n$ is the feasible set of codewords according to the input cost constraint. Rearranging~\eqref{kappa-1} then yields 
\begin{align}
&P_{Y^n|X^n=x^n}\left[\bigcup_{j=1}^{M} \tilde{i}(x^n(j);Y^n)>\log \gamma_n\right]  > \epsilon -P_{Y^n|X^n=x^n}[\,\tilde{i}(x^n;Y^n)\leq\log \gamma_n] \geq \tau^* \label{union}
\end{align}
again for all sequences $x^n\in\mathcal{F}_n$, where
\begin{align}
\tau^*=\epsilon- \sup_{x^n\in F}P_{Y^n|X^n=x^n}[\,\tilde{i}(x^n;Y^n)\leq \log \gamma_n]. \label{tau*}
\end{align} 
Now, thinking of the union in the brackets in~\eqref{union} as a binary test on $Y^n$, one can cast the problem into the framework of the following composite hypothesis test, which is used to treat the input cost constraint:
\begin{align}
&\kappa_{\tau}\left(\{P_{Y^n|X^n=x^n}\}_{x^n\in F}, Q_{Y^n}\right) := \min_{Z: P_{Y^n|X^n=x^n}[Z(Y^n)=1]>\tau, \forall x^n\in F} Q_{Y^n}[Z(Y^n)=1],
\end{align}
where $Z(Y^n)$ is a binary test choosing either the class of conditional channel laws~$\{P_{Y^n|X^n=x^n}\}_{x^n\in F}$ if $Z=1$, or the unconditional output distribution~$Q_{Y^n}$ if $Z=0$. 
The $\kappa\beta$ bound of~\cite{PPV} for maximal error probability can then be stated as
\begin{align}
\kappa_{\tau^*}\left(\{P_{Y^n|X^n=x^n}\}_{x^n\in F}, Q_{Y^n}\right) &\leq Q_{Y^n}\left[\bigcup_{j=1}^{M}\tilde{i}(x^n(j);Y^n)>\log \gamma_n\right] \\
&  \leq M \sup_{x^n\in F} Q_{Y^n}[\,\tilde{i}(x^n;Y^n)>\log \gamma_n]. \label{kappa-2}
\end{align}

Interpretation of the composite hypothesis test~$\kappa_\tau$ and accordingly its evaluation for the P2P Gaussian channel is quite involved. Polyanskiy \emph{et al.}~\cite{PPV} invoke arguments from abstract algebra to analyze the performance of this test for the feasible set $\mathcal{F}_n=\{x^n\in\mathbb{R}^n: ||x^n||=\sqrt{nP}\}$ being the ``power shell'' and the special choice $Q_{Y^n}\sim\mathcal{N}(\mathbf{0},(1+P)I_n)$ with the selection~$\tau^*=1/\sqrt{n}$ in~\eqref{tau*}, finally concluding that
\begin{align}
\log\kappa_{\tau^*}\geq \frac{1}{2}\log n + O(1), \label{kappa-bound}
\end{align}
which with application of the CLT results in the following second-order optimal achievable rate for the P2P Gaussian channel
\begin{align}
\frac{\log M}{n}\leq C(P) - \sqrt{\frac{V(P)}{n}}Q^{-1}(\epsilon) + O\left(\frac{1}{n}\right),
\end{align}
where~$V(P)$ is the dispersion of the Gaussian P2P channel
\begin{align}
V(P)&=\frac{\log^2e}{2}\frac{P(P+2)}{(1+P)^2}. 
\label{dispersion}
\end{align}

Comparing the $\kappa\beta$ bound of~\cite{PPV} with the random coding and typicality decoding method discussed earlier suggests an important insight. Introducing the composite hypothesis bound $\kappa_\tau$ in~\cite{PPV} enables a \emph{change of measure} from $P_{Y^n|X^n=x^n}$ in~\eqref{kappa-1} to $Q_{Y^n}$ in~\eqref{kappa-2} in computing the confusion probability. A similar process occurs in the random coding argument with typicality decoding, as the random generation of the codebook makes it possible to change the measure for computation of the confusion probability from $P_{Y^n|X^n=x^n}$ in~\eqref{typ-1} to its average $P_{Y^n}$ in~\eqref{typ-2}. We suspect the reason why the composite $\kappa_\tau$ test is introduced in~\cite{PPV} is to enable such a change of measure argument which is required for the evaluation of the confusion probability, but is not directly available in the sequential generation of maximal coding, which does not incorporate any random generation process. This insight is one of the main ideas we will use in this paper for the analysis of the P2P Gaussian channel and the Gaussian MAC with a random coding and typicality decoding method.

\subsection{Shannon's Geometric Bound}
\label{sub-Shannon}

As mentioned before, the best known achievable rate for P2P Gaussian channel is due to Shannon~\cite{Shannon} who starts with a random codebook generation according to the uniform distribution on the $n$-dimensional sphere of radius~$\sqrt{nP}$, i.e. the power shell
\begin{align}
||x^n||^2=nP, \label{pshell}
\end{align}
but considers the optimal ML decoding method. Since this rule is equivalent to minimum Euclidian distance in~$\mathbb{R}^n$, Shannon employs geometric arguments to evaluate and bound the code-ensemble-average probability that the i.i.d. Gaussian channel noise moves the output closer to some incorrect codeword than to the originally transmitted codeword:
\begin{align}
\epsilon
&=-\int_0^\pi \left\{1-\left[1-\frac{S_n(1;\theta)}{S_n(1)}\right]^{M-1}\right\}dQ(\theta) \\
&\leq Q(\theta^*)-\frac{M}{S_n(1)}\int_0^{\theta^*} S_n(1;\theta)dQ(\theta), 
\end{align}
where: $S_n(1;\theta)$ is the surface area of a unit-radius $n$-dimensional spherical cap with half-angle $\theta$; $S_n(1)=S_n(1;\pi)$ is the surface area of a unit-radius $n$-dimensional sphere; $Q(\theta)$ is the probability with respect to $\mathcal{N}(\mathbf{0},I_n)$ that a point $x^n\in\mathbb{R}^n$ with $||x^n||=\sqrt{nP}$ is moved outside a circular cone of half-angle $\theta$ with vertex at the origin and axis passing through $x^n$; and $\theta^*$ is a characteristic of the rate defined as the solid angle satisfying $S_n(1;\theta^*)=S_n(1)/M$. Shannon then expresses this geometric bound as an error exponent result in terms of the rate and SNR: 
\begin{align}
\epsilon \leq \frac{\alpha(P,\theta^*)}{\sqrt{n}} e^{-nE(P,\theta^*)}
\end{align}
where $\alpha(P,\theta^*)$ and $E(P,\theta^*)$ are positive functions of the power~$P$ and the rate characteristic~$\theta^*$.

A key observation in Shannon's work is his use of the uniform distribution on the power shell, which enables him to develop sharp non-asymptotic bounds. In this paper, we will follow Shannon in this respect, but rely on the more familiar and less complex method of typicality decoding which we show is still capable of achieving sharp non-asymptotic bounds for the Gaussian channel, at least up to the second order.

Having reviewed the basic elements of the different procedures for handling cost constraints, especially in Gaussian settings, we now move on to the formal statement of our problems and results.

\section{P2P Gaussian Channel}
\label{sec-P2P}

In this section, we re-derive the fundamental communication limits over the P2P Gaussian channel, that are well-known from classical and recent studies, e.g.~\cite{Shannon, PPV, Hayashi}, and were summarized in Section II. Building upon the standard random coding and typicality decoding method with slight modifications, our aims are both to 1) clarify our proof techniques in this simpler setting before exploring the more complex Gaussian MAC model, and 2) provide a more transparent alternative achievability proof for the P2P Gaussian problem, which is at least second-order optimal.

\subsection{System Model and Known Result}
\label{sub-P2P-model}

A general P2P channel with input cost constraint and without feedback consists of an input alphabet $\mathcal{X}$, an output alphabet $\mathcal{Y}$, and an $n$-letter channel transition probability $P_{Y^n|X^n}(y^n|x^n)\colon\mathcal{F}_n\to\mathcal{Y}^n$, where $\mathcal{F}_n\subseteq\mathcal{X}^n$ is the feasible set of $n$-letter input sequences. 
For such a P2P channel, an $(n,M,\epsilon)$ code is composed of a message set~$\mathcal{M}=\{1,...,M\}$ and a corresponding set of codewords and mutually exclusive decoding regions $\{(x^n(j),D_{j})\}$ with $j\in\mathcal{M}$, such that the average error probability satisfies
\begin{align}
P_e^{(n)}:=\frac{1}{M}\sum_{j=1}^{M} \Pr[Y^n\notin D_{j}|X^n(j)\;\text{sent}]\leq\epsilon. \label{err-p2p}
\end{align}
Accordingly, a rate $\frac{\log M}{n}$ is \emph{achievable} for the P2P channel with finite blocklength~$n$ and average error probability~$\epsilon$ if such an $(n,M,\epsilon)$ code exists. 

In particular, a P2P memoryless Gaussian channel without feedback consists of an input and an output taking values on the real line~$\mathbb{R}$ and a channel transition probability density $P_{Y|X}(y|x)\colon\mathbb{R}\to\mathbb{R}$ whose $n$-th extension follows $\mathcal{N}(y^n; x^n,I_n)$, i.e.,
\begin{align}
P_{Y^n|X^n}(y^n|x^n)\!=\!\prod_{t=1}^n\!P_{Y|X}(y_t|x_t)\!=\!(2\pi)^{-n/2} e^{-||y^n-x^n||^2/2}\!.
\end{align}
For such a P2P Gaussian channel, an $(n,M,\epsilon,P)$ code is an $(n,M,\epsilon)$ code as defined above, in which each codeword also satisfies a maximal power constraint:
\begin{align}
\frac{1}{n}\sum_{t=1}^n x_t^2(j)=\frac{1}{n}||x^n(j)||^2\leq P, \qquad \forall j\in\mathcal{M}.
\end{align}
Accordingly, a rate $\frac{\log M}{n}$ is \emph{achievable} for the P2P Gaussian channel with finite blocklength~$n$, average error probability $\epsilon$, and maximal power~$P$ if such an $(n,M,\epsilon,P)$ code exists. 

The set of all achievable second-order coding rates for the P2P Gaussian channel is characterized as~\cite{PPV,Hayashi}
\begin{align}
\frac{\log M}{n}\leq C(P) - \sqrt{\frac{V(P)}{n}}Q^{-1}(\epsilon) + O\left(\frac{1}{n}\right), \label{2nd-P2P}
\end{align}
where $C(P)$ and $V(P)$ are the capacity~\eqref{capacity} and dispersion~\eqref{dispersion} of the P2P Gaussian channel, respectively. This section presents a relatively straight-forward achievability proof for this result based upon random coding and typicality decoding.

\subsection{Key Elements of the Proof}
\label{sub-key-p2p}

In this section, we summarize the main ingredients of the proof of main result~\eqref{2nd-P2P} for P2P Gaussian channels. The formal proof will be given in the Sections~\ref{sub-P2P-Ach} and~\ref{sub-2nd-p2p}.  The three main ingredients are modified random coding and typicality decoding, CLT for functions of random vectors, and change of measure and uniform bounding. 

\subsubsection{Modified Random Coding and Typicality Decoding}
\label{sub-P2P-RC-TD} 

The random coding and typicality decoding bounds~\eqref{typ-2} and~\eqref{DT-cost-PPV} are by now the most standard method for proving the achievability side of the channel coding theorems~\cite{ElGamal-Kim}. If the input distribution were chosen to be i.i.d., such as the i.i.d. Gaussian distribution, then an evaluation of these achievability bounds would be straightforward, using a CLT for the outage probability, and an LDT bound for the confusion probability, but the cost-violation probability would be non-zero $P_{X^n}[X^n\notin\mathcal{F}_n]\neq0$. However, as discussed in Section I, for the P2P Gaussian channel (and potentially other cost-constrained channels), no single-letter i.i.d. input distribution exists which can achieve the second-order optimal performance~\eqref{2nd-P2P}, and more complicated $n$-letter input distributions must be considered. A non-i.i.d. input distribution~$P_{X^n}$ that is second-order optimal and leads to a zero cost-violation probability~$P_{X^n}[X^n\notin\mathcal{F}_n]=0$, such as the uniform distribution on the power shell~\eqref{pshell}, induces a non-i.i.d. output distribution~$P_{Y^n}$, and this in turn prevents the mutual information RV~$i(X^n;Y^n)$ from being a sum of independent random variables, i.e. $i(X^n;Y^n)\neq\sum_{t=1}^n i(X_t;Y_t)$, a form which is convenient for CLT and LDT analyses. It is therefore appealing for typicality decoding to change the reference of the mutual information RV from the actual output distribution~$P_{Y^n}$ to an arbitrary \textit{product} distribution~$Q_{Y^n}$ and work with a \emph{modified} mutual information RV~$\tilde{i}(X^n;Y^n)$ which is defined as 
\begin{equation}
\tilde{i}(X^n;Y^n):=\log \frac{P_{Y^n|X^n}(Y^n|X^n)}{Q_{Y^n}(Y^n)}, \label{mod-i}
\end{equation}
 and can be written as a summation $\tilde{i}(X^n;Y^n)=\sum_{t=1}^n \tilde{i}(X_t;Y_t)$, although the summands are not independent.

\subsubsection{CLT for Functions of Random Vectors}
\label{sub-CLT-Taylor}

The second-order approximations of channel coding rates mainly result from approximating the mutual information RV in the outage probability with a Gaussian distribution via the CLT. In the conventional setting, the CLT applies to a summation of independent RV's. However, due to the use of non-i.i.d. input distribution to handle the cost constraint, in this paper, we deal with mutual information densities that are sums of non-independent random variables or vectors. Rather, they can be expressed as (vector-) \emph{functions of sums} of i.i.d. random vectors. To facilitate the CLT for these situations, we rely on a simplified version of a technical result of Hoeffding and Robbins~\cite[Theorem 4]{Hoef}, for which they also credit Anderson and Rubin~\cite{Rubin}. Since these references do not specify the rate of convergence to Gaussianity, we slightly extend the analysis to prove a Berry-Esseen version of their result. The basic idea of the proof, which is relegated to Appendix A, is the application of Taylor's Theorem around the mean to a (vector-) function whose arguments are normalized sums of i.i.d. random vectors.

\begin{proposition} \label{CLT-func}
Let $\{\mathbf{U}_t:=(U_{1t},...,U_{Kt})\}_{t=1}^\infty$ be zero-mean i.i.d. random vectors in $\mathbb{R}^K$ with $\mathbb{E}[||\mathbf{U}_{1}||_2^3]<\infty$, and denoting $\mathbf{u}:=(u_1,...,u_K)$, let $\mathbf{f}(\mathbf{u}):\mathbb{R}^K\to\mathbb{R}^L$ be an $L$-component vector-function~$\mathbf{f}(\mathbf{u})=(f_1(\mathbf{u}),...,f_L(\mathbf{u}))$ which has continuous second-order partial derivatives in a $K$-hypercube neighborhood of $\mathbf{u}=\mathbf{0}$ of side length at least~$\frac{1}{\sqrt[4]{n}}$, and whose corresponding Jacobian matrix~$\mathbf{J}$ at $\mathbf{u}=\mathbf{0}$ consists of the following first-order partial derivatives
\begin{align}
J_{lk}:=\left.\frac{\partial f_l(\mathbf{u})}{\partial u_k}\right|_{\mathbf{u}=\mathbf{0}} \qquad l=1,...,L, \quad k=1,...,K.
\end{align}
Then, for any convex Borel-measurable set $\mathcal{D}$ in $\mathbb{R}^L$, there exists a finite positive constant~$B$ such that\footnote{The gap to Gaussianity in this form of CLT, similar to other Berry-Esseen type bounds, is on the order of $\frac{1}{\sqrt{n}}$. Although this is enough for second-order proofs, it makes the reader doubt whether this slowly decaying gap leads to accurate approximations for short blocklengths. This is indeed a valid concern, but one should note that empirical evidences, such as the results of~\cite{PPV} for discrete and Gaussian P2P channels, suggest that CLT is actually a highly accurate estimate and that the Berry-Esseen bound may be (much) looser than reality; c.f.~\cite[P. 135]{P-thesis}.}
\begin{align}
\left| \Pr\left[\mathbf{f}\left(\frac{1}{n}\sum_{t=1}^n \mathbf{U}_t\right) \in \mathcal{D}\right] - \Pr\left[\mathcal{N}\left(\mathbf{f}\left(\mathbf{0}\right), \mathbf{V}\right) \in \mathcal{D}\right] \right| \leq \frac{B}{\sqrt{n}}, 
\end{align}
where the covariance matrix~$\mathbf{V}$ is given by $\mathbf{V}=\frac{1}{n}\mathbf{J}\text{Cov}(\mathbf{U}_1)\mathbf{J}^T$, that is, its entries are defined as
\begin{align}
V_{ls}:= \frac{1}{n}\sum_{k=1}^K \sum_{p=1}^K J_{lk} J_{sp} \mathbb{E}[U_{k1}U_{p1}], \qquad l,s=1,...,L.
\end{align}
\end{proposition}

We would like to mention that, references~\cite{PPV,Hayashi} take an indirect approach based on symmetry to handle this problem for the P2P Gaussian channel, thus reducing the problem to the evaluation of the conditional outage probability for a fixed input sequence, for which the conventional CLT is applicable. However, this approach does not generalize to multi-user settings. Our approach here to use a CLT for functions of random variables, although more complicated, provides a direct analysis of outage probability without exploiting the symmetry property, and will be seen to generalize to the Gaussian MAC.

\subsubsection{Change of Measure and Uniform Bounding}
\label{sub-P2P-change-measure}

For non-i.i.d. input distributions~$P_{X^n}$, such as the uniform distribution on the power shell~\eqref{pshell}, an LDT analysis of the confusion probability is also challenging due to the non-product nature of the output distribution~$P_{Y^n}$ induced by the input distribution~$P_{X^n}$. We are therefore interested in changing the measure with respect to which the confusion probability is analyzed, as follows:
\begin{align}
&P_{X^n}P_{Y^n}[\,\tilde{i}(X^n;Y^n)>\log \gamma_n]  \label{chng-beg} \\
&= \int\int 1\left\{\tilde{i}(x^n;y^n)>\log \gamma_n\right\} dP_{Y^n}(y^n) dP_{X^n}(x^n)  \\
&=\int\int1\left\{\tilde{i}(x^n;y^n)>\log\gamma_n\right\}\frac{dP_{Y^n}(y^n)}{dQ_{Y^n}(y^n)}dQ_{Y^n}(y^n)dP_{X^n}(x^n)   \\
&\leq \sup_{y^n\in\mathcal{Y}^n} \frac{dP_{Y^n}(y^n)}{dQ_{Y^n}(y^n)}\, P_{X^n}Q_{Y^n}\!\!\left[\,\tilde{i}(X^n;Y^n)>\log\gamma_n\right]. \label{change-measure}
\end{align}
The final expression~\eqref{change-measure} enables us to compute the confusion probability with respect to the more convenient measure~$Q_{Y^n}$, but at the expense of the additional Radon-Nikodym (R-N) derivative~$\frac{dP_{Y^n}(y^n)}{dQ_{Y^n}(y^n)}$.\footnote{To be precise, the definition of this R-N derivative requires an absolute continuity condition $P_{Y^n}\ll Q_{Y^n}$~\cite{R-N-der}, which is considered to be true for our general arguments in Theorem~\ref{corol-modified-DT}, and can be easily seen to hold in our concrete example of the P2P Gaussian channel.} This bound would be particularly useful, if this extra coefficient is uniformly bounded by a positive constant~$K$ or a slowly growing function~$K_n$, such that its rate loss does not affect the second-order behavior. 

A close examination of~\cite{PPV} shows that the $\kappa_\tau$ performance characteristic in the $\kappa\beta$ bound is also mainly concerned with the R-N derivative~$\frac{dP_{Y^n}(Y^n)}{dQ_{Y^n}(Y^n)}$ introduced above, and the bound~\eqref{kappa-bound} is analogous to the uniform bounding by~$K_n$ in the analysis above~\eqref{change-measure}. We believe the difference is that our analysis using random coding, typicality decoding, and change of measure is a more transparent procedure and more closely follows conventional lines of argument.

\vspace{3mm}

We are now ready to provide the formal proof in the next two subsections.

\subsection{Non-Asymptotic Achievability for Cost-Constrained Channels}
\label{sub-P2P-Ach}

In the following, we state a result based upon modified random coding and typicality decoding for achievability on general P2P channels with input cost constraints valid for any blocklength. The result basically describes the error probability in terms of the outage, confusion, and constraint-violation probabilities, and is based on the dependence testing (DT) bound of \cite{PPV}.

\begin{theorem}
\label{corol-modified-DT}
For a general P2P channel $(\mathcal{X},P_{Y^n|X^n}(y^n|x^n),\mathcal{Y})$, any input distribution $P_{X^n}$, and any output distribution~$Q_{Y^n}$, there exists an $(n,M,\epsilon)$ code that satisfys the input cost constraint~$\mathcal{F}_n$ and
\begin{align}
\epsilon \leq P_{X^n}P_{Y^n|X^n}[\,\tilde{i}(X^n;Y^n)\leq \log \gamma_n] + K_n\frac{M-1}{2} P_{X^n}Q_{Y^n}[\,\tilde{i}(X^n;Y^n)>\log \gamma_n] + P_{X^n}[X^n\!\!\notin\mathcal{F}_n], \label{DT-simple}
\end{align}
where the coefficient $K_n$ is defined as
\begin{align}
K_n:=\sup_{y^n\in\mathcal{Y}^n} \frac{dP_{Y^n}(y^n)}{dQ_{Y^n}(y^n)}, \label{unif-bound}
\end{align}
and $\gamma_n$ is an arbitrary positive threshold whose optimal choice to give the highest rates is $\gamma_n\equiv K_n\frac{M-1}{2}$.
\end{theorem}

\textit{Remark.} The bound~\eqref{DT-simple} reduces to a standard one with random coding, typicality decoding, and $K_n=1$ if the auxiliary distribution~$Q_{Y^n}(y^n)$ is identical to the actual output distribution~$P_{Y^n}(y^n)$ induced by the input~$P_{X^n}(x^n)$. 

\begin{proof}
The channel encoder randomly generates $M$~codewords of the codebook independently according to some given $n$-letter distribution $P_{X^n}$, where $n$ is the designated blocklength. Observing the output $y^n$, the decoder chooses the first codeword $x^n(\hat{m})$ of the codebook which looks ``typical'' with $y^n$ in a \emph{modified} one-sided sense 
\begin{equation}
\tilde{i}(x^n(\hat{m});y^n)>\log \gamma(x^n(\hat{m})), \label{mod-decod}
\end{equation}
where $\gamma(x^n)$ is a codeword-dependent threshold and $\tilde{i}(x^n;y^n)$ is the corresponding realization of the \emph{modified} mutual information random variable~$\tilde{i}(X^n;Y^n)$.
The error probability averaged over the set of $M$ codewords of all possible realizations of the codebook can then be bounded, similar to~\eqref{typ-1}-\eqref{typ-2}, as the sum of an outage probability and a confusion probability as follows:
\begin{align}
\epsilon &\leq P_{X^n}P_{Y^n|X^n}[\,\tilde{i}(X^n;Y^n)\leq \log \gamma(X^n)] + \frac{M-1}{2} P_{X^n}P_{Y^n}[\,\tilde{i}(X^n;Y^n)>\log \gamma(X^n)] . 
\end{align}
Applying the change of measure technique of~\eqref{change-measure} with the definition~\eqref{unif-bound} yields 
\begin{align}
\epsilon &\leq P_{X^n}P_{Y^n|X^n}[\,\tilde{i}(X^n;Y^n)\leq \log \gamma(X^n)] + K_n\frac{M-1}{2} P_{X^n}Q_{Y^n}[\,\tilde{i}(X^n;Y^n)>\log \gamma(X^n)] . 
\end{align}
Upon selecting the threshold~$\log\gamma(x^n)$ as in~\eqref{inf-th} and following the reasonings proceeding~\eqref{DT-cost-1}-\eqref{DT-cost-PPV} to handle the cost constraint, we infer that there exists a deterministic codebook, consisting of $M$ codewords all belonging to the feasible set~$\mathcal{F}_n$, whose average error probability~$\epsilon$ satisfies
\begin{align}
\epsilon \leq P_{X^n}[X^n\notin\mathcal{F}_n] + P_{X^n}P_{Y^n|X^n}[\,\tilde{i}(X^n;Y^n)\leq \log \gamma_n] + K_n\frac{M-1}{2} P_{X^n}Q_{Y^n}[\,\tilde{i}(X^n;Y^n)>\log \gamma_n] . \label{typ-2-mod}
\end{align}

To conclude the final assertion of Theorem~\ref{corol-modified-DT}, it is sufficient to observe that the last two summands on the RHS of \eqref{typ-2-mod} are a weighted sum of two types of error in a Bayesian binary hypothesis test, and therefore corresponds to average error probability of the test. Then, it is known from Neyman-Pearson Theorem that the optimal test is a likelihood-ratio test (LRT), as we have used in~\eqref{mod-decod}, with the optimal threshold equal to the ratio of priors or simply the ratio of the coefficients of the two error probabilities of the test, namely $\gamma_n\equiv K_n\frac{M-1}{2}$.
\end{proof}

\subsection{Second-Order Characterization for P2P Gaussian Channels}
\label{sub-2nd-p2p}

So far, we have stated and proved Theorem 1 which holds for any arbitrary cost-constrained P2P channel. In the following, we specialize this achievability bound to the P2P Gaussian channel.

\subsubsection{Coding on the Power Shell}

First, we choose the input distribution as the uniform distribution on the power shell:
\begin{align}
P_{X^n}(x^n)=\frac{\delta(||x^n||-\sqrt{nP})}{S_n(\sqrt{nP})}, \label{unif-shell}
\end{align}
where $\delta(\cdot)$ is the Dirac delta function and~$S_n(r)=\frac{2\pi^{n/2}}{\Gamma(n/2)}r^{n-1}$ is the surface area of an~$n$-dimensional sphere of radius~$r$. Notice that this distribution satisfies the input power constraint with probability one, so that
\begin{align}
P_{X^n}[X^n\notin\mathcal{F}_n]=P_{X^n}[||X^n||^2>nP]=0. 
\label{cost=0}
\end{align}
Moreover, the output distribution induced by this input is
\begin{align}
P_{Y^n}(y^n) & =\int_{\mathbb{R}^n} P_{X^n}(x^n) P_{Y^n|X^n}(y^n|x^n) dx^n \\
& = \int_{\mathbb{R}^n} \frac{\delta(||x^n||-\sqrt{nP})}{S_n(\sqrt{nP})} (2\pi)^{-n/2} e^{-||y^n-x^n||^2/2} dx^n \\
& = \int_0^\pi \int_{0}^\infty \frac{\delta(r-\sqrt{nP})}{S_n(\sqrt{nP})} (2\pi)^{-n/2}e^{-r^2/2}   e^{-||y^n||^2/2}e^{||y^n||r\cos\theta} S_{n-1}(r\sin\theta) r dr d\theta \label{decomp} \\
& = \frac{(2\pi)^{-n/2}\Gamma\left(n/2\right)}{\pi^{1/2}\Gamma\left(\frac{n-1}{2}\right)} e^{-nP/2} e^{-||y^n||^2/2}   \int_0^\pi  e^{||y^n||\sqrt{nP}\cos\theta} \sin^{n-2}\theta d\theta \\
& = \frac{(2\pi)^{-n/2}\Gamma\left(n/2\right)}{\pi^{1/2}\Gamma\left(\frac{n-1}{2}\right)} e^{-nP/2} e^{-||y^n||^2/2}    \frac{\pi^{1/2}2^{n/2-1}\Gamma\left(\frac{n-1}{2}\right)}{(||y^n||\sqrt{nP})^{n/2-1}}I_{n/2-1}(||y^n||\sqrt{nP}) \label{bessel} \\
& = \frac{1}{2}\pi^{-n/2}\Gamma\left(\frac{n}{2}\right) e^{-nP/2} e^{-||y^n||^2/2}  \frac{I_{n/2-1}(||y^n||\sqrt{nP})}{(||y^n||\sqrt{nP})^{n/2-1}},  \label{shell-output}
\end{align}
where~\eqref{decomp} follows form decomposing the space~$\mathbb{R}^n$ into a continuum of $(n-1)$-dimensional ring elements of radius $r\sin\theta$ with $0\leq r\leq \infty$ being the distance of ring points from the origin and $0\leq\theta\leq\pi$ being the angle of ring points with the line connecting the origin and the point~$y^n$, and where~\eqref{bessel} follows from the definition of modified Bessel function~$I_v(\cdot)$ of the first kind and~$v$-th order. It is worth mentioning that the general form of the above marginal distribution is obtained in~\cite{Bessel}. 

Next, we select the reference output distribution for the P2P Gaussian channel as
\begin{align}
Q_{Y^n}(y^n)=\mathcal{N}(y^n;\mathbf{0},(1+P)I_n), \label{Q-p2p}
\end{align}
that is, the capacity-achieving output distribution. 
The following proposition will then bound the R-N derivative term introduced in~\eqref{unif-bound}. The proof, which is a slight generalization of that in~\cite[p.~2347]{PPV}, is given in Appendix~\ref{proof-Prop-P2P}.

\begin{proposition}
\label{Prop-P2P-unif-bound}
Let $P_{Y^n}$ be the distribution~\eqref{shell-output} induced on the output of the P2P Gaussian channel by the uniform input distribution~\eqref{unif-shell} on the power shell, and let $Q_{Y^n}$ be the capacity-achieving output distribution~\eqref{Q-p2p} for the P2P Gaussian channel. There exists a positive constant~$K$ such that, for sufficiently large~$n$,
\begin{align}
\frac{dP_{Y^n}(y^n)}{dQ_{Y^n}(y^n)}\leq K, \qquad \forall\; y^n\in\mathbb{R}^n; 
\end{align}
In fact, $K\leq1$ is a constant independent of the power constraint $P$.
\end{proposition}
\textit{Remark.} Using some more complicated manipulations, this proposition can be shown to be valid for any finite~$n$, but the above statement is enough for our second-order analysis.

Proposition~\ref{Prop-P2P-unif-bound} facilitates the use of Theorem~\ref{corol-modified-DT} with the aforementioned choices for the input distribution~$P_{X^n}$ and the reference output distribution~$Q_{Y^n}$. Substituting~\eqref{cost=0} into the achievability bound~\eqref{DT-simple} of Theorem~\ref{corol-modified-DT} with the optimal threshold~$\gamma_n\equiv K\frac{M-1}{2}$ only leaves the outage and confusion probabilities.  
In the following, we evaluate the outage and confusion probabilities for sufficiently large blocklength to obtain second-order achievable rates.

\subsubsection{Evaluation of the Outage Probability}
\label{sub-P2P-outage}

In this subsection, we bound the outage probability 
\begin{align}
P_{X^n}P_{Y^n|X^n}\left[\,\tilde{i}(X^n;Y^n)\leq \log\gamma_n\right]  
\end{align}
where the input distribution $P_{X^n}$ is the uniform distribution on the power shell~\eqref{unif-shell}. Note that, since the input distribution is non-i.i.d., the summands in $\tilde{i}(X^n;Y^n)=\sum_{t=1}^n \tilde{i}(X_t;Y_t)$ are not independent so that direct application of the conventional CLT is not possible. Unlike the indirect symmetry-based approach of~\cite{PPV,Hayashi}, we here give a direct, although more complicated, analysis of the outage probability which does not rely on the conditional mutual information RV and instead makes use of the structure of the uniform distribution on the power shell. 

Under the $P_{Y^n|X^n}$ law, the output~$Y^n$ can be written in the form
\begin{align}
Y^n=X^n+Z^n,
\end{align}
where $Z^n\sim \mathcal{N}(\mathbf{0},I_n)$ is the i.i.d. unit-variance channel noise. With the choice~\eqref{Q-p2p} for $Q_{Y^n}(y^n)$, the modified mutual information random variable simplifies as
\begin{align}
\tilde{i}(X^n;Y^n)&\equiv \log\frac{(2\pi)^{-n/2}e^{-||Y^n-X^n||^2/2}}{(2\pi(1+P))^{-n/2}e^{-||Y^n||^2/2(1+P)}} \\
&=\frac{n}{2}\log(1+P)+\frac{\log e}{2}\left[\frac{||Y^n||^2}{1+P}-||Y^n-X^n||^2\right] \\
&=nC(P)+\frac{\log e}{2(1+P)}\left[||X^n+Z^n||^2-(1+P)||Z^n||^2\right] \\
&=nC(P) +\frac{\log e}{2(1+P)}\left[P(n-||Z^n||^2)+2 \langle X^n,Z^n \rangle\right]. \label{mi-p2p}
\end{align}
where~\eqref{mi-p2p} uses the inner-product notation $\langle a^n,b^n\rangle:=\sum_{t=1}^n a_t b_t$ and the fact that $||X^n||^2=nP$ with probability one. 

Although this random variable is written in the form of a summation, the summands are not independent, since the input $X^n$ is not independent across time. However, recall that independent uniform RVs on the power shell are functions of independent Gaussian RVs. More precisely, let $W^n\sim\mathcal{N}(\mathbf{0},I_n)$ be an i.i.d. Gaussian RV independent of the noise RV~$Z^n$. The input elements $X_t$, $t=1,...,n$, of the independent uniformly distributed RV $X^n$ on the power shell can then be expressed as follows:
\begin{align}
X_t=\sqrt{nP}\frac{W_t}{||W^n||}.
\end{align}

To apply the CLT for functions of Proposition~\ref{CLT-func}, consider the sequence $\{\mathbf{U}_t:=(U_{1t},U_{2t},U_{3t})\}_{t=1}^\infty$ whose elements are defined as
\begin{align}
U_{1t}&=1-Z_t^2, \\
U_{2t}&=\sqrt{P}W_tZ_{t}, \\
U_{3t}&=W_t^2-1.
\end{align}
Note that this random vector has an i.i.d. distribution across time $t=1,...,n$, and its moments can be easily verified to satisfy $\mathbb{E}[\mathbf{U}_1]=0$ and $\mathbb{E}[||\mathbf{U}_t||_2^3]<\infty$. Moreover, the covariance matrix of this vector is given by
\begin{align}
\text{Cov}(\mathbf{U})=
\begin{bmatrix}
2 & 0    & 0 \\
0        & P    & 0  \\
0        & 0    & 2 
\end{bmatrix}.
\end{align}

Next, define the function $f$ as
\begin{align}
f(\mathbf{u})&=Pu_1+ \frac{2u_2}{\sqrt{1+u_3}}.
\end{align}
Notice that, $f(\mathbf{0})=0$ and all the first- and second-order partial derivatives of $f$ are continuous in a neighborhood of $\mathbf{u}=\mathbf{0}$. Moreover, the Jacobian matrix $\{\frac{\partial f(\mathbf{u})}{\partial u_j}\}_{1\times3}$ at $\mathbf{u}=\mathbf{0}$ can be readily verified to be
\begin{align}
\left.J\right|_{\mathbf{u}=\mathbf{0}}=
\begin{bmatrix}
P & 2 & 0
\end{bmatrix}.
\end{align}
We are therefore left with 
\begin{align}
f\left(\frac{1}{n}\sum_{t=1}^n \mathbf{U}_t\right)
&=\frac{P}{n}\sum_{t=1}^n (1-Z_t^2)+ \frac{2\frac{1}{n}\sum_{t=1}^n \sqrt{P}W_tZ_t}{\sqrt{1+\frac{1}{n}\sum_{t=1}^n (W_t^2-1)}} \label{f-sum-1} \\
&= \frac{1}{n}\sum_{t=1}^n P(1-Z_t^2)+ \frac{2}{n}\sum_{t=1}^n  \frac{\sqrt{nP}W_t}{||W^n||}Z_t \label{f-sum-2} \\
&= \frac{1}{n}\left[P(n-||Z^n||^2)+2 \langle X^n,Z^n \rangle \right]. \label{f-sum-3}
\end{align}

We now conclude from Proposition~\ref{CLT-func} that the modified mutual information RV~\eqref{mi-p2p} converges in distribution to a Gaussian distribution with mean $nC(P)$ and variance given by  
\begin{align}
\left(\frac{n\log e}{2(1+P)}\right)^2\frac{1}{n}
\begin{bmatrix}
P & 2 & 0
\end{bmatrix}
\begin{bmatrix}
2 & 0    & 0 \\
0        & P    & 0  \\
0        & 0    & 2 
\end{bmatrix}
\begin{bmatrix}
P \\ 2 \\ 0
\end{bmatrix} 
= n \frac{\log^2 e}{2}\frac{P(P+2)}{(1+P)^2}=nV(P).
\end{align}
In particular, the outage probability is bounded as
\begin{align}
P_{X^n}P_{Y^n|X^n}\left[\,\tilde{i}(X^n;Y^n) \leq \log\left(K\frac{M-1}{2}\right)\right] 
&\leq \Pr\left[\mathcal{N}(nC(P),nV(P)) \leq \log\left(K\frac{M-1}{2}\right)\right] +\frac{B_1}{\sqrt{n}} \\
&= Q\left(\frac{nC(P)-\log\left(K\frac{M-1}{2}\right)}{\sqrt{nV(P)}}\right) + \frac{B_1}{\sqrt{n}}, \label{outage-P2P}
\end{align}
where $B_1$ is the constant introduced in Proposition~\ref{CLT-func}.

\subsubsection{Evaluation of the Confusion Probability}
\label{sub-2nd-Ach}

In this subsection, we bound the confusion probability 
\begin{align}
K\frac{M-1}{2}P_{X^n}Q_{Y^n}\left[\,\tilde{i}(X^n;Y^n)\leq \log\gamma_n\right]  
\end{align}
where the input distribution $P_{X^n}$ is the uniform distribution on the power shell~\eqref{unif-shell}, and $Q_{Y^n}$ is the capacity-achieving output distribution~\eqref{Q-p2p}. We first need a change of measure technique as in~\cite{PPV}
\begin{align}
Q\left[\frac{dP}{dQ}>\gamma\right]&=\int1\left\{\frac{dP}{dQ}>\gamma\right\}dQ \\
&=\int \left(\frac{dP}{dQ}\right)^{-1}1\left\{\frac{dP}{dQ}>\gamma\right\}dP \\
&=\mathbb{E}_{P}\left[\left(\frac{dP}{dQ}\right)^{-1}1\left\{\frac{dP}{dQ}>\gamma\right\}\right]. \label{change of measure}
\end{align}
Using~\eqref{change of measure} with $P_{Y^n|X^n=x^n}$ in the role of $P$ and~$Q_{Y^n}$ in the role of~$Q$, we can bound the conditional confusion probability, conditioned on the input sequence~$x^n$ on the power shell, as follows:
\begin{align}
&Q_{Y^n}\left[\tilde{i}(x^n;Y^n)\!>\!\log\left(K\frac{M-1}{2}\right)\right] \notag \\
&=\mathbb{E}_{P_{Y^n|X^n=x^n}}\left[\exp\{-\tilde{i}(x^n;Y^n)\}1\left\{\tilde{i}(x^n;Y^n)\!>\!\log\left(K\frac{M-1}{2}\right)\right\}\right] \\
&=\mathbb{E}_{P_{Y^n|X^n=x^n}}\left[\exp\left\{-\sum_{t=1}^n \tilde{i}(x_t;Y_t)\right\}1\left\{\sum_{l=1}^n \tilde{i}(x_t;Y_t)>\log\left(K\frac{M-1}{2}\right)\right\}\right] \\
&\leq \frac{B_{2}}{\sqrt{n}}\left(K\frac{M-1}{2}\right)^{-1}, \label{PPV-lemma20}
\end{align}
where \eqref{PPV-lemma20} is a refined large deviation bound according to~\cite[Lemma~47]{PPV}. The specific expression for the finite constant~$B_2$ can be computed readily in terms of the power constraint~$P$, but is not necessary here. Since the bound~\eqref{PPV-lemma20} is uniform with respect to the actual input sequence~$x^n$, the unconditional confusion probability can be bounded as
\begin{align}
K\frac{M-1}{2} P_{X^n}Q_{Y^n}\left[\tilde{i}(X^n;Y^n)\!>\!\log\left(K\frac{M-1}{2}\right)\right] \leq \frac{B_{2}}{\sqrt{n}}.
\label{conf-P2P}
\end{align}

\subsubsection{Completion}

Substituting~\eqref{cost=0},~\eqref{outage-P2P}, and~\eqref{conf-P2P} into the achievability bound~\eqref{DT-simple} of Theorem~\ref{corol-modified-DT} and recalling~\eqref{err-p2p} that, with a little abuse of notation cf.~\cite[Eq. (186)]{PPV}, $\epsilon$ is the target error probability, yields 
\begin{align}
\epsilon \geq Q\left(\frac{nC(P)-\log\left(K\frac{M-1}{2}\right)}{\sqrt{nV(P)}}\right) + \frac{{B}}{\sqrt{n}},
\end{align}
where~${B}=B_1+B_2$. Upon rearranging we obtain
\begin{align}
\log M &\leq nC(P)-\sqrt{nV(P)}Q^{-1}\left(\epsilon- \frac{{B}}{\sqrt{n}}\right) - \underbrace{\log K}_{= O(1)} \\
&= nC(P)-\sqrt{nV(P)}Q^{-1}(\epsilon) + \sqrt{nV(P)}  O\left(\frac{1}{\sqrt{n}}\right) + O(1)  \label{Taylor} \\
 &= nC(P)-\sqrt{nV(P)}Q^{-1}(\epsilon) + O(1),
\end{align}
where~\eqref{Taylor} follows from the Taylor expansion for the $Q^{-1}$ function
\begin{align}
Q^{-1}\left(\epsilon- \frac{{B}}{\sqrt{n}}\right)=Q^{-1}(\epsilon) + \underbrace{\left.\frac{dQ^{-1}(x)}{dx}\right|_{x=\epsilon}}_{= O(1)} \frac{{B}}{\sqrt{n}} + o(\frac{1}{\sqrt{n}}) 
= Q^{-1}(\epsilon) + O\left(\frac{1}{\sqrt{n}}\right). 
\end{align}

Thus, we have proved that an $(n,M,\epsilon,P)$ code exists if the rate satisfies
\begin{align}
\frac{\log M}{n} &\leq C(P)-\sqrt{\frac{V(P)}{n}} Q^{-1}(\epsilon) + O\left(\frac{1}{n}\right),
\label{P2P-ach-approx}
\end{align}
where $C(P)$ and $V(P)$ are the capacity and dispersion of the P2P Gaussian channel, respectively. 
As discussed in Section I, we have observed that such high rates arise from coding schemes in which outages dominate confusions.

\section{Gaussian MAC}
\label{sec-MAC}

In this section, we study our main problem of interest, namely the fundamental communication limits over the Gaussian MAC in the finite blocklength regime. We first state our main result on achievable second-order coding rate regions for the Gaussian MAC, overview the key elements of the proof, and then develop the results in detail.

\subsection{System Model and Main Results}
\label{sub-MAC-model}

A general 2-user multiple access channel (MAC) with input cost constraints and without feedback consists of two input alphabets $\mathcal{X}_1$ and $\mathcal{X}_2$, an output alphabet~$\mathcal{Y}$, and an $n$-letter channel transition probability given by $P_{Y^n|X_1^nX_2^n}(y^n|x_1^n,x_2^n)\colon\mathcal{F}_{1n}\times\mathcal{F}_{2n}\to\mathcal{Y}^n$, where $\mathcal{F}_{1n}\subseteq\mathcal{X}_1^n$ and $\mathcal{F}_{2n}\subseteq\mathcal{X}_2^n$ are the feasible sets of $n$-letter input sequences for the two users, respectively.
For such a MAC, an $(n,M_1,M_2,\epsilon)$ code is composed of two message sets $\mathcal{M}_1=\{1,...,M_1\}$ and $\mathcal{M}_2=\{1,...,M_2\}$, and a corresponding set of codeword pairs and mutually exclusive decoding regions $\{(x_1^n(j),x_2^n(k),D_{j,k})\}$, with $j\in\mathcal{M}_1$ and $k\in\mathcal{M}_2$, such that the average error probability satisfies
\begin{align}
P_e^{(n)}:=\frac{1}{M_1M_2}\sum_{j=1}^{M_1}\sum_{k=1}^{M_2} \Pr[Y^n\notin D_{j,k}|X_1^n(j),X_2^n(k)\;\text{sent}]\leq\epsilon. \label{err-mac}
\end{align}
Accordingly, a $\left(\frac{\log M_1}{n},\frac{\log M_2}{n}\right)$ rate pair is \emph{achievable} for this MAC with finite blocklength $n$ and average error probability $\epsilon$ if such an $(n,M_1,M_2,\epsilon)$ code exists.

In particular, a memoryless 2-user Gaussian  MAC without feedback consists of two inputs and an output all taking values on the real line~$\mathbb{R}$ and a channel transition probability density $P_{Y|X_1X_2}(y|x_1,x_2)\colon\mathbb{R}\times\mathbb{R}\to\mathbb{R}$ whose $n$-th extension follows $\mathcal{N}(y^n; x_1^n+x_2^n,I_n)$, i.e.,
\begin{align}
P_{Y^n|X_1^nX_2^n}(y^n|x_1^n,x_2^n)&=\prod_{t=1}^n P_{Y|X_1X_2}(y_t|x_{1t},x_{2t})=(2\pi)^{-n/2} e^{-||y^n-x_1^n-x_2^n||^2/2}.
\end{align}
For such a Gaussian MAC, an $(n,M_1,M_2,\epsilon,P_1,P_2)$ code is an $(n,M_1,M_2,\epsilon)$ code as defined above, in which each codeword also satisfies a maximal power constraint:
\begin{align}
\frac{1}{n}\sum_{t=1}^n x_{1t}^2(j)=\frac{1}{n}||x_1^n(j)||^2&\leq P_1, \qquad \forall j\in\mathcal{M}_1,\\
\frac{1}{n}\sum_{t=1}^n x_{2t}^2(k)=\frac{1}{n}||x_2^n(k)||^2&\leq P_2, \qquad \forall k\in\mathcal{M}_2.
\end{align}
Accordingly, a rate pair $\left(\frac{\log M_1}{n},\frac{\log M_2}{n}\right)$ is \emph{achievable} for the Gaussian MAC with finite blocklength $n$, average error probability $\epsilon$, and maximal power constraints~$P_1$ and $P_2$ if such an $(n,M_1,M_2,\epsilon,P_1,P_2)$ code exists.

From classical results by Cover~\cite{Cover-GMAC} and Wyner~\cite{Wyner}, we know that the capacity region of the Gaussian MAC in the infinite blocklength regime, that is when $n\to\infty$, is given by the pentagonal region
\begin{align}
\frac{\log M_1}{n} &\leq C(P_1) + o(1), \\
\frac{\log M_2}{n} &\leq C(P_2) + o(1), \\
\frac{\log M_1}{n} + \frac{\log M_2}{n} &\leq C(P_1+P_2) + o(1).
\end{align}
Some old and new bounds on the error exponent of the Gaussian MAC are also known, e.g.~\cite{Gallager-MAC,Pradhan,Haim-EE}. 
In the following, we state finite-blocklength achievability results that lead to the following achievable second-order rate regions for the Gaussian MAC. As mentioned in Section I,   the following regions are achieved using codebooks that are randomly generated according to independent uniform distributions on the respective power shells. 
\begin{theorem}
\label{GMAC-joint}
\textit{(Joint Outage)} An achievable region for the 2-user Gaussian MAC with maximal power constraints~$P_1$ and~$P_2$ is given by the union of all rate pairs $(\frac{\log M_1}{n},\frac{\log M_2}{n})$ satisfying
\begin{align}
\begin{bmatrix}
\frac{\log M_1}{n}\\
\frac{\log M_2}{n}\\
\frac{\log M_1}{n}\!+\!\frac{\log M_2}{n}
\end{bmatrix}
\in\mathbf{C}(P_1,P_2)-\frac{1}{\sqrt{n}} Q^{-1}(\epsilon;\mathbf{V}(P_1,P_2))+O\left(\frac{1}{n}\right)\mathbf{1},
\end{align}
where: $\mathbf{1}=\begin{bmatrix} 1 & 1 & 1\end{bmatrix}^T$ denotes the all-one vector; $Q^{-1}(\epsilon;\mathbf{\Sigma})$ is the inverse complementary CDF of a 3-dimensional Gaussian random variable defined as the set
\begin{align}
Q^{-1}(\epsilon;\mathbf{\Sigma})\!:=\!\left\{\mathbf{z}\in\mathbb{R}^3: \Pr\left(\mathcal{N}(\mathbf{0},\mathbf{\Sigma})\leq\mathbf{z}\right)\geq1-\epsilon\right\},
\label{Q-inv}
\end{align}
with vector-inequality understood element-wise; the capacity vector $\mathbf{C}(P_1,P_2)$ and dispersion matrix $\mathbf{V}(P_1,P_2)$ are defined as
\begin{align}
\mathbf{C}(P_1,P_2):=
\begin{bmatrix}
C(P_1)\\
C(P_2)\\
C(P_1+P_2)
\end{bmatrix},
\label{capacity vector}
\end{align}
and
\begin{align}
\mathbf{V}(P_1,P_2):=
\begin{bmatrix}
V(P_1)&V_{1,2}(P_1,P_2)&V_{1,3}(P_1,P_2) \\
V_{1,2}(P_1,P_2)&V(P_2)&V_{2,3}(P_1,P_2) \\
V_{1,3}(P_1,P_2)&V_{2,3}(P_1,P_2)&V(P_1+P_2)+ V_{3}(P_1,P_2)
\end{bmatrix}\!\!,
\label{dispersion matrix}
\end{align}
in which $C(P)$ and $V(P)$ are the capacity and dispersion of the P2P Gaussian channel, respectively, 
\begin{align}
C(P)&=\frac{1}{2}\log(1+P),  \\
V(P)&=\frac{\log^2e}{2}\frac{P(P+2)}{(1+P)^2},
\end{align}
and we have employed the shorthands
\begin{align}
V_{1,2}(P_1,P_2) &= \frac{\log^2 e}{2}\frac{P_1P_2}{(1+P_1)(1+P_2)},  \label{cross-disp-1}  \\
V_{u,3}(P_1,P_2) &= \frac{\log^2 e}{2}\frac{P_u(2+P_1+P_2)}{(1+P_u)(1+P_1+P_2)}, \quad u\in\{1,2\}\label{cross-disp-2} \\
V_{3}(P_1,P_2) & = \log^2 e\frac{P_1P_2}{(1+P_1+P_2)^2}.  \label{disp-3}  
\end{align}
\end{theorem}

The evaluation of the above region, especially when extended to a large number of users, may be cumbersome. In the following, we present another second-order achievable rate region, which is easier to compute even for large number of users, but as we have already seen in Figure 1 provides a very good estimate of the joint-outage region for the Gaussian MAC.

\begin{theorem}
\label{GMAC-splitting}
\textit{(Outage Splitting)}  An achievable region for the 2-user Gaussian MAC with maximal power constraints~$P_1$ and~$P_2$ is given by the union of all $(\frac{\log M_1}{n},\frac{\log M_2}{n})$ pairs satisfying
\begin{align}
&\frac{\log M_1}{n}\leq C(P_1)-\sqrt{\frac{V(P_1)}{n}}Q^{-1}(\lambda_1\epsilon)\!+\!O\left(\frac{1}{n}\right), \notag\\ 
&\frac{\log M_2}{n}\leq C(P_2) -\sqrt{\frac{V(P_2)}{n}}Q^{-1}(\lambda_2\epsilon)\!+\!O\left(\frac{1}{n}\right), \notag\\ 
&\frac{\log M_1}{n}+\frac{\log M_2}{n}\leq C(P_1+P_2)-\sqrt{\frac{V(P_1+P_2)+ V_{3}(P_1,P_2)}{n}}Q^{-1}(\lambda_3\epsilon)+O\left(\frac{1}{n}\right), \label{b-3}
\end{align}
for some choice of positive constants $\lambda_1,\lambda_2,\lambda_3$ satisfying $\lambda_1+\lambda_2+\lambda_3=1$.
\end{theorem}

Both of the achievable second-order rate regions in Theorems~\ref{GMAC-joint} and~\ref{GMAC-splitting} suggest that taking finite blocklength into account introduces a rate penalty (for the interesting case of $\epsilon<\frac{1}{2}$) that depends on blocklength, error probability, and Gaussian MAC dispersions. 
However, the main difference between the two theorems is that, in Theorem~\ref{GMAC-splitting}, the average error probability~$\epsilon$ is basically split among the three outage events of a 2-user Gaussian MAC according to some $(\lambda_1,\lambda_2,\lambda_3)$ partitioning and a one-dimensional CLT is applied. A similar approach was taken in~\cite{ML-ISIT12,Verdu-Allerton12} for the MAC in the discrete setting. On the other hand, in Theorem~\ref{GMAC-joint}, essentially all the average error probability~$\epsilon$ is assigned to the joint outage event and a multi-dimensional CLT is applied. This latter approach, which leads to a relatively larger region, is similar to that in~\cite{Tan,HM, Albert} for the discrete MAC. Finally, we would like to point out that the statements of Theorems 2 and 3 correct a slight error in the corresponding result in our conference version of this work~\cite{ML-Allerton12}, in which the term $V_3(P_1,P_2)$ defined in~\eqref{disp-3} was missing in~\eqref{dispersion matrix} and~\eqref{b-3}.

To observe the tightness of the region achieved by random codebooks with independent power shell input distribution, we compare it with several other second-order inner and outer rate regions relying on simple and common structures. First, consider the second-order rate region achieved by a pair of random codebooks which are, as usual~\cite{Cover}, generated according to independent i.i.d. Gaussian distributions. One can easily show an extension of~\eqref{Gaussian-input} to a Gaussian MAC so that
\begin{align}
\begin{bmatrix}
\frac{\log M_1}{n}\\
\frac{\log M_2}{n}\\
\frac{\log M_1}{n}\!+\!\frac{\log M_2}{n}
\end{bmatrix}&\in\mathbf{C}(\bar{P}_1,\bar{P}_2)-\frac{1}{\sqrt{n}}Q^{-1}\left(\epsilon; \mathbf{V}_{\text{G}}(\bar{P}_1,\bar{P}_2)\right)+O\left(\frac{1}{n}\right)\mathbf{1},
\label{Gaussian-input-MAC}
\end{align}
are achievable, where $\bar{P}_1\!=\!P_1\!-\!\delta$ and $\bar{P}_2\!=\!P_2\!-\!\delta$ for an arbitrarily small positive constant\footnote{Again, the margin~$\delta$ can be vanishing with blocklength provided that it decays strictly slower than $O\left(\frac{1}{\sqrt{n}}\right)$.}~$\delta$, and 
\begin{align}
\mathbf{V}_{\text{G}}(P_1,P_2)=\log^2 e
\begin{bmatrix}
\frac{P_1}{1+P_1}\!&\!\frac{P_1P_2}{2(1+P_1)(1+P_2)}\!\!&\!\!\frac{P_1(2+2P_1+P_2)}{2(1+P_1)(1+P_1+P_2)} \\
\frac{P_1P_2}{2(1+P_1)(1+P_2)}\!&\!\frac{P_2}{1+P_2}\!\!&\!\!\frac{P_2(2+P_1+2P_2)}{2(1+P_2)(1+P_1+P_2)}  \\
\frac{P_1(2+2P_1+P_2)}{2(1+P_1)(1+P_1+P_2)}\!&\!\frac{P_2(2+P_1+2P_2)}{2(1+P_2)(1+P_1+P_2)}\!\!&\!\!\frac{P_1+P_2}{1+P_1+P_2}
\end{bmatrix}.
\end{align} 

Another important comparison is with the rate region achieved by a pair of independent truncated Gaussian random codebooks, as employed by Gallager for the error exponent analysis of the Gaussian MAC~\cite{Gallager-MAC}. This rate region is given by the set of all rate pairs $(R_1:=\frac{\log M_1}{n},R_2:=\frac{\log M_1}{n})$ satisfying
\begin{align}
\epsilon \leq a n 2^{-nE_1\left(R_1\right)} +a n 2^{-nE_2\left(R_2\right)} + a n^2 2^{-nE_3\left({R_1+R_2}\right)},
\end{align}
where $a$ is a constant, and the error exponent term for the individual rates is defined as
\begin{align}
E_l(R_l):= 
\begin{dcases}
\frac{(P_l-\alpha_l)\log e}{2^{2R_l+1}}+ \frac{1}{2}\log\left(2^{2R_l}-\alpha_l\right) & \text{if} \quad \frac{1}{2}\log\left(\frac{2+P_l+\sqrt{4+P_l^2}}{4}\right) \leq R_l \leq C(P_l) \\
\left(1-\beta_l+\frac{P_l}{2}\right)\log e+\frac{1}{2}\log\left(\beta_l\left[\beta_l-\frac{P_l}{2}\right]\right)-R_l & \text{if} \quad 0 \leq R_l < \frac{1}{2}\log\left(\frac{2+P_l+\sqrt{4+P_l^2}}{4}\right)
\end{dcases}
\end{align}
with $l=1,2$ and the shorthands
\begin{align}
\alpha_l &:= \frac{P_l(2^{2R_l}-1)}{2}\left[\sqrt{1+\frac{2^{2R_l+2}}{P_l(2^{2R_l}-1)}}-1\right], \\
\beta_l &:= \frac{1}{2} \left[1+\frac{P_l}{2}+\sqrt{1+\frac{P_l^2}{2}}\,\right];
\end{align}
and the error exponent term for the sum-rate is defined as 
\begin{align}
E_3(R_s):= 
\begin{dcases}
(1+\rho-\theta_1)\log e + \log\left(\frac{\theta_1}{1+\rho}\right) \\
\qquad \qquad  \qquad \qquad \qquad \text{if} \;\; \frac{1}{2}\log\left(\frac{1}{2}\left[1-\frac{P_s}{4}+\sqrt{1-\frac{P_s}{2}+\frac{P_s^2}{4}}\,\right]\right) \leq R_s \leq C(P_s) \\
2\left[\log\left(\frac{\theta_2}{2}\right)-(\theta_2+1)\log e\right]+\frac{1}{2}\log\left(1+\frac{P_s}{\theta_2}\right)-R_s  \\
\qquad \qquad \qquad \qquad \qquad \text{if} \;\; 0 \leq R_s < \frac{1}{2}\log\left(\frac{1}{2}\left[1-\frac{P_s}{4}+\sqrt{1-\frac{P_s}{2}+\frac{P_s^2}{4}}\,\right]\right)
\end{dcases}
\end{align}
with the shorthands~$R_s:=R_1+R_2$, $P_s:=P_1+P_2$ and
\begin{align}
\rho&:=  \left[\frac{1}{2}+\frac{2^{2R_s+1}}{P_s}-\frac{1}{2}\sqrt{1+\frac{2^{2R_s+3}}{P_s}+\frac{2^{2R_s+4}}{P_s^2}}\,\right]^{-1/2}-1, \\
\theta_1 &:= \frac{1+\rho-P_s}{2} +\frac{1}{2}\sqrt{P_s^2+2P_s+(1+\rho)^2}, \\
\theta_2 &:=1-\frac{P_s}{2}+\frac{1}{2} \log\left(P_s^2+2P_s+4\right).
\end{align}

It is also interesting to compare with the second-order achievable region via time-division multiple access~(TDMA). For TDMA with power control, the two users can share the~$n$ channel uses, use single-user coding strategies, and average the error probability~$\epsilon$. Specifically, user 1 transmits in the first~$\alpha n$ channel uses with power~$P_1/\alpha$ and rate such that an average error probability~$\beta\epsilon$ is achieved, and user 2 transmits in the remaining~$\bar{\alpha}n:=(1-\alpha)n$ channel uses with power~$P_2/\bar{\alpha}$ and rate such that an average error probability~$\tilde{\beta}\epsilon$ is achieved. Since the average error probability of this scheme is $\epsilon=\beta\epsilon+\tilde{\beta}\epsilon-\beta\tilde{\beta}\epsilon^2$, we choose $\tilde{\beta}=(1-\beta)/(1-\beta\epsilon)$. Using the power shell uniform input distribution for each user and relying on the Gaussian P2P results~\cite{PPV,Hayashi}, the TDMA strategy achieves the following set of rate pairs:
\begin{align}
\frac{\log M_1}{n}&\leq\alpha C\left(\frac{P_1}{\alpha}\right)\!-\!\sqrt{\frac{\alpha}{n} V\left(\frac{P_1}{\alpha}\right)}Q^{-1}(\beta\epsilon)+O\left(\frac{1}{n}\right), \notag\\ 
\frac{\log M_2}{n}&\leq\bar{\alpha} C\left(\frac{P_2}{\bar{\alpha}}\right)-\sqrt{\frac{\bar{\alpha}}{n}V\left(\frac{P_2}{\bar{\alpha}}\right)}Q^{-1}\left(\frac{(1-\beta)\epsilon}{1-\beta\epsilon}\right)+O\left(\frac{1}{n}\right),  
\end{align}
for some $0\leq\alpha\leq1$ and $0\leq\beta\leq1$.

Further comparison can be made using single-user outer bounds. Since the achievable rate for each user cannot exceed that when the other user is silent, similar to~\cite{Tan}, two simple outer bounds can be developed using single-user results~\cite{PPV,Hayashi} by assigning the total error probability~$\epsilon$ to only one of the outage events. Hence
\begin{align}
&\frac{\log M_1}{n}\leq C(P_1)-\sqrt{\frac{V(P_1)}{n}}Q^{-1}(\epsilon)\!+\!O\left(\frac{\log n}{n}\right), \\ 
&\frac{\log M_2}{n}\leq C(P_2) -\sqrt{\frac{V(P_2)}{n}}Q^{-1}(\epsilon)\!+\!O\left(\frac{\log n}{n}\right), 
\end{align}
presents a simple outer bound for the Gaussian MAC. Note that, since the two power constraints $||x_1^n||^2\leq nP_1$ and $||x_2^n||^2\leq nP_2$ do not imply the sum-power constraint $||x_1^n+x_2^n||^2\leq n(P_1+P_2)$, a similar conclusion cannot be readily made for the sum-rate, that is, the inequality
\begin{align}
\frac{\log M_1}{n}+\frac{\log M_2}{n}\leq C(P_1+P_2)-\sqrt{\frac{V(P_1+P_2)}{n}}Q^{-1}(\epsilon)+O\left(\frac{\log n}{n}\right)
\end{align}
is not a trivial outer bound, as was mistakenly claimed in~\cite{ML-Allerton12}; however, it is conjectured to be a valid outer bound.

Finally, we would like to mention a \textit{hypothetical} second-order rate region which would be achievable if the sum of power shell inputs fell on the \textit{sum-power shell}, i.e., if the two users' codebooks were independent and distributed uniformly on the respective power shells \textit{and} the equality $||x_1^n(j)+x_2^n(k)||^2=||x_1^n(j)||^2+||x_2^n(k)||^2$ hypothetically hold for (almost) all codeword pairs. Following the lines of proof of Theorem 2, such a hypothetical codebook pair would then achieve the following second-order rate region:
\begin{align}
\begin{bmatrix}
\frac{\log M_1}{n}\\
\frac{\log M_2}{n}\\
\frac{\log M_1}{n}+\frac{\log M_2}{n}
\end{bmatrix}
\in\mathbf{C}(P_1,P_2)-\frac{1}{\sqrt{n}} Q^{-1}(\epsilon;\mathbf{V}_{\text{sum}}(P_1,P_2))+O\left(\frac{1}{n}\right)\mathbf{1}, \label{sum-power-shell}
\end{align}
where $\mathbf{V}_{\text{sum}}(P_1,P_2)$ is defined as the rank-2 matrix 
\begin{align}
\mathbf{V}_{\text{sum}}(P_1,P_2)=
\begin{bmatrix}
V(P_1)&V_{1,2}(P_1,P_2)&V_{1,3}(P_1,P_2) \\
V_{1,2}(P_1,P_2)&V(P_2)&V_{2,3}(P_1,P_2) \\
V_{1,3}(P_1,P_2)&V_{2,3}(P_1,P_2)&V(P_1+P_2)
\end{bmatrix}\!\!,
\label{Vout}
\end{align}
and all other notations are defined as in Theorem 2. Note that the only difference between the above region and that stated in Theorem 2 is the term $V_3(P_1,P_2)$ in the last diagonal element of the dispersion matrix~$\mathbf{V}(P_1,P_2)$ in~\eqref{dispersion matrix} which is dropped in $\mathbf{V}_{\text{sum}}(P_1,P_2)$. The term~$V_3(P_1,P_2)$ captures the variance of the inner product $\langle X_1^n,X_2^n \rangle$, which is the remainder term in $||x_1^n+x_2^n||^2-||x_1^n||^2-||x_2^n||^2=2\langle x_1^n,x_2^n \rangle$. Since $V_3(P_1,P_2)$ is a positive term, its removal leads to a ``smaller'' dispersion matrix and hence a lower rate penalty. In fact, as was illustrated in Figure~1, one can show that the power-shell rate region of Theorem 2 is roughly \textit{halfway} between the i.i.d. Gaussian rate region and this hypothetical sum-power shell region. A similar observation has been made by Gallager in the study of error exponents for the Gaussian MAC~\cite{Gallager-MAC}. We conjecture that the hypothetical sum-power shell rate region provides a second-order \textit{outer} region for the Gaussian MAC, with the third-order term in~\eqref{sum-power-shell} replaced by~$O\left(\frac{\log n}{n}\right)\mathbf{1}$.  

A numerical comparison of these different rate regions was already presented in Figure 1 of Section I. The reminder of this section presents a relatively straightforward proof of Theorems 2 and 3 based upon random coding and typicality decoding as well as the application of the CLT for functions.

\subsection{Key Elements of the Proof}
\label{sub-key-mac}

In this section, we comment on the main ingredients of the proof of Theorems 2 and 3 for the Gaussian MAC. The details of the proofs will be given in Sections~\ref{sub-MAC-Ach} and~\ref{sub-2nd-mac}.

\subsubsection{Modified Random Coding and Typicality Decoding}

Analogously to the P2P Gaussian channel, we show that i.i.d. input distributions are not sufficient to achieve the second-order optimal performance over the Gaussian MAC. However, the analysis of the conventional achievability result based upon random coding and  typicality decoding for the MAC is difficult for non-i.i.d. inputs, particularly because 1) the induced (conditional) output distributions $P_{Y^n|X_2^n}, P_{Y^n|X_1^n}, P_{Y^n}$ are not i.i.d. and 2) the corresponding mutual information RVs cannot be written as a sums. Hence, we use \textit{modified} mutual information RVs defined in terms of arbitrary reference output distributions~$Q^{(1)}_{Y^n|X_2^n}, Q^{(2)}_{Y^n|X_1^n}, Q^{(3)}_{Y^n}$, instead of the actual output distributions:
\begin{subequations}
\begin{align}
\tilde{i}(X_1^n;Y^n|X_2^n)&:=\log\frac{P_{Y^n|X_1^nX_2^n}(Y^n|X_1^n,X_2^n)}{Q^{(1)}_{Y^n|X_2^n}(Y^n|X_2^n)}, \\
\tilde{i}(X_2^n;Y^n|X_1^n)&:=\log\frac{P_{Y^n|X_1^nX_2^n}(Y^n|X_1^n,X_2^n)}{Q^{(2)}_{Y^n|X_1^n}(Y^n|X_1^n)}, \\
\tilde{i}(X_1^nX_2^n;Y^n)&:=\log\frac{P_{Y^n|X_1^nX_2^n}(Y^n|X_1^n,X_2^n)}{Q^{(3)}_{Y^n}(Y^n)};
\end{align}
\label{mod-mi-MAC}%
\end{subequations}
Selecting the reference distributions to be in product from, e.g. $Q^{(1)}_{Y^n|X_2^n}(y^n|x_2^n)=\prod_{t=1}^n Q^{(1)}_{Y_t|X_{2t}}(y_t|x_{2t})$, enables us to write these RVs as sums of random variables, e.g. $\tilde{i}(X_1^n;Y^n|X_2^n)=\sum_{t=1}^n \tilde{i}(X_{1t};Y_t|X_{2t})$, a form which is convenient for the later application of CLT and LDT. However we note that these summands are not independent.

\subsubsection{CLT for Functions of Random Vectors}

As mentioned earlier, the summands of the modified mutual information random variables of a Gaussian MAC are not independent. Moreover, the interaction of the two users' codebooks through the inner product~$\langle X_1^n,X_2^n \rangle$ prevents the application of symmetry arguments as used in~\cite{PPV,Hayashi} for the P2P Gaussian channel. However, these mutual information RVs can be expressed as (vector-) functions of i.i.d. random vectors, to which the CLT for functions in Propositon~\ref{CLT-func} can be applied. Specifically, this proposition with $L=3$ will be used in the the proof of the joint-outage region in Theorem 2 and with $L=1$ in the proof of the outage-splitting region in Theorem 3. 

\subsubsection{Change of Measure and Uniform Bounding}

Similar to the P2P case, an LDT analysis of the three confusion probabilities in the modified random coding and typicality decoding bounds is challenging due to the non-product nature of the (conditional) output distributions induced by a pair of non-i.i.d. input distributions. Thus, we again apply a change of measure argument for computing these confusion probabilities. Analogously to~\eqref{chng-beg}-\eqref{change-measure}, we can show that the following inequalities hold:
\allowdisplaybreaks{
\begin{subequations}
\begin{align}
&P_{X_1^n}P_{X_2^n}P_{Y^n|X_2^n}\left[\,\tilde{i}(X_1^n;Y^n|X_2^n)> \log\gamma_1(X_1^n,X_2^n)\right] \notag \\
& \qquad \qquad \leq \sup_{x_2^n\in\mathcal{X}_2^n,\,y^n\in\mathcal{Y}^n}\frac{dP_{Y^n|X_2^n}(y^n|x_2^n)}{dQ^{(1)}_{Y^n|X_2^n}(y^n|x_2^n)} \, P_{X_1^n}P_{X_2^n}Q^{(1)}_{Y^n|X_2^n}\left[\,\tilde{i}(X_1^n;Y^n|X_2^n)> \log\gamma_1(X_1^n,X_2^n)\right]\, \\
&P_{X_1^n}P_{X_2^n}P_{Y^n|X_1^n}\left[\,\tilde{i}(X_2^n;Y^n|X_1^n)> \log\gamma_2(X_1^n,X_2^n)\right] \notag \\
&\qquad \qquad \leq \sup_{x_1^n\in\mathcal{X}_2^n, \,y^n\in\mathcal{Y}^n}\frac{dP_{Y^n|X_1^n}(y^n|x_1^n)}{dQ^{(2)}_{Y^n|X_1^n}(y^n|x_1^n)} \, P_{X_1^n}P_{X_2^n}Q^{(2)}_{Y^n|X_1^n}\left[\,\tilde{i}(X_2^n;Y^n|X_1^n)> \log\gamma_2(X_1^n,X_2^n)\right]\, \\
&P_{X_1^n}P_{X_2^n}P_{Y^n}\left[\,\tilde{i}(X_1^nX_2^n;Y^n)> \log\gamma_3(X_1^n,X_2^n)\right] \notag \\
&\qquad \qquad \leq \sup_{y^n\in\mathcal{Y}^n}\frac{dP_{Y^n}(y^n)}{dQ^{(3)}_{Y^n}(y^n)} \, P_{X_1^n}P_{X_2^n}Q^{(3)}_{Y^n}\left[\,\tilde{i}(X_1^nX_2^n;Y^n)> \log\gamma_3(X_1^n,X_2^n)\right]. 
\end{align}
\label{change-measure-MAC}%
\end{subequations}}%
Therefore, we may compute the confusion probabilities with respect to the more convenient measures~$Q^{(1)}_{Y^n|X_2^n}$, $Q^{(2)}_{Y^n|X_1^n}$, $Q^{(3)}_{Y^n}$, but at the expense of the additional R-N derivatives.\footnote{Again, the absolute continuity conditions for the above R-N derivatives are implicitly assumed to hold in the general bounds of Theorem~\ref{corol-modified-DT-MAC} and can be easily verified for the Gaussian MAC.} The bounds in~\eqref{change-measure-MAC} are particularly useful if these extra coefficients can bounded by positive constant~$K_1,K_2,K_3$ or slowly growing functions~$K_{1n},K_{2n},K_{3n}$, such that their rate loss does not affect the second-order behavior.

\vspace{3mm}
We are now ready to provide the formal proof in the next two subsections.

\subsection{Non-Asymptotic Achievability for Cost-Constrained MAC}
\label{sub-MAC-Ach}

In the following, we state a result based upon random coding and modified typicality decoding for achievability on a general MAC with input cost constraints valid for any blocklength. The result basically describes the error probability in terms of the outage, confusion, and constraint-violation probabilities.

\begin{theorem}
\label{corol-modified-DT-MAC}
For a MAC $(\mathcal{X}_1,\mathcal{X}_2,P_{Y^n|X_1^nX_2^n}(y^n|x_1^n,x_2^n),\mathcal{Y})$, for any pair of independent input distributions~$P_{X_1^n}$ and~$P_{X_2^n}$, and any triple of (conditional) output distributions $Q^{(1)}_{Y^n|X_2^n}, Q^{(2)}_{Y^n|X_1^n}, Q^{(3)}_{Y^n}$, there exists an $(n,M_1,M_2,\epsilon)$ code satisfying input cost constraints~$\mathcal{F}_{1n}$ and~$\mathcal{F}_{2n}$ with
\begin{align}
\epsilon \leq\;  & P_{X_1^n}P_{X_2^n}P_{Y^n|X_1^nX_2^n}\left[\,\tilde{i}(X_1^n;Y^n|X_2^n)\leq \log\gamma_{1n} \right. \notag \\
& \qquad \qquad \qquad \qquad \quad \cup\left. \tilde{i}(X_2^n;Y^n|X_1^n)\leq \log\gamma_{2n} \right. \notag \\
& \qquad \qquad \qquad \qquad \quad \cup\left. \tilde{i}(X_1^nX_2^n;Y^n)\leq \log\gamma_{3n}\,\right] \notag\\
& +K_{1n}\frac{M_1-1}{2}P_{X_1^n}P_{X_2^n}Q^{(1)}_{Y^n|X_2^n}\left[\,\tilde{i}(X_1^n;Y^n|X_2^n)>\log\gamma_{1n}\right] \notag\\
& +K_{2n}\frac{M_2-1}{2}P_{X_1^n}P_{X_2^n}Q^{(2)}_{Y^n|X_1^n}\left[\,\tilde{i}(X_2^n;Y^n|X_1^n)>\log\gamma_{2n}\right] \notag\\
& +K_{3n}\frac{(M_1-1)(M_2-1)}{2}P_{X_1^n}P_{X_2^n}Q^{(3)}_{Y^n}\left[\,\tilde{i}(X_1^nX_2^n;Y^n)>\log\gamma_{3n}\right] \notag \\
& + P_{X_1^n}P_{X_2^n}[X_1^n\notin\mathcal{F}_{1n} \cup X_2^n\notin\mathcal{F}_{2n}], 
\label{DT-joint}
\end{align}
or
\begin{align}
\epsilon \leq\; & P_{X_1^n}P_{X_2^n}P_{Y^n|X_1^nX_2^n}\left[\,\tilde{i}(X_1^n;Y^n|X_2^n)\leq \log\gamma_{1n}\right] \notag\\
&+K_{1n}\frac{M_1-1}{2}P_{X_1^n}P_{X_2^n}Q^{(1)}_{Y^n|X_2^n}\left[\,\tilde{i}(X_1^n;Y^n|X_2^n)>\log\gamma_{1n}\right] \notag\\
&+ P_{X_1^n}P_{X_2^n}P_{Y^n|X_1^nX_2^n}\left[\,\tilde{i}(X_2^n;Y^n|X_1^n)\leq \log\gamma_{2n}\right] \notag\\
&+K_{2n}\frac{M_2-1}{2}P_{X_1^n}P_{X_2^n}Q^{(2)}_{Y^n|X_1^n}\left[\,\tilde{i}(X_2^n;Y^n|X_1^n)>\log\gamma_{2n}\right] \notag\\
&+ P_{X_1^n}P_{X_2^n}P_{Y^n|X_1^nX_2^n}\left[\,\tilde{i}(X_1^nX_2^n;Y^n)\leq \log\gamma_{3n}\right] \notag\\
&+K_{3n}\frac{(M_1-1)(M_2-1)}{2}P_{X_1^n}P_{X_2^n}Q^{(3)}_{Y^n}\left[\,\tilde{i}(X_1^nX_2^n;Y^n)>\log\gamma_{3n}\right] \notag \\
&+ P_{X_1^n}P_{X_2^n}[X_1^n\notin\mathcal{F}_{1n} \cup X_2^n\notin\mathcal{F}_{2n}], 
\label{DT-splitting}
\end{align}
where: the modified mutual information random variables for the MAC are defined in~\eqref{mod-mi-MAC}; the coefficients~$K_{1n}$, $K_{2n}$, $K_{3n}$ are defined as\footnote{The bounds~\eqref{DT-joint} and~\eqref{DT-splitting} can be further improved by replacing the $\sup_{x_2^n\in\mathcal{X}_2^n,\,y^n\in\mathcal{Y}^n}$ in~\eqref{u-1} with $\sup_{x_2^n\in\text{supp}(P_{X_2^n}),\,y^n\in\mathcal{Y}^n}$, where $\text{supp}(P_{X_2^n})$ denotes the support of the distribution~$P_{X_2^n}$, and analogously in~\eqref{u-2}, but is not necessary in this work.}
\begin{subequations}
\begin{align}
K_{1n}&:=\sup_{x_2^n\in\mathcal{X}_2^n,\,y^n\in\mathcal{Y}^n}\frac{dP_{Y^n|X_2^n}(y^n|x_2^n)}{dQ^{(1)}_{Y^n|X_2^n}(y^n|x_2^n)}, \label{u-1}\\
K_{2n}&:=\sup_{x_1^n\in\mathcal{X}_2^n, \,y^n\in\mathcal{Y}^n}\frac{dP_{Y^n|X_1^n}(y^n|x_1^n)}{dQ^{(2)}_{Y^n|X_1^n}(y^n|x_1^n)}, \label{u-2} \\
K_{3n}&:=\sup_{y^n\in\mathcal{Y}^n}\frac{dP_{Y^n}(y^n)}{dQ^{(3)}_{Y^n}(y^n)};
\end{align}
\label{unif-bound-MAC}%
\end{subequations}
and $\gamma_{1n},\gamma_{2n},\gamma_{3n}$ are arbitrary positive thresholds whose optimal choices to give the highest rates in~\eqref{DT-splitting} are
\begin{align}
\gamma_{1n}\equiv K_{1n}\frac{M_1-1}{2}, \qquad \gamma_{2n}\equiv K_{2n}\frac{M_2-1}{2}, \qquad \gamma_{3n}\equiv K_{3n}\frac{(M_1-1)(M_2-1)}{2}\;. \label{th-MAC}
\end{align}
\end{theorem}

The achievable bounds~\eqref{DT-joint} and~\eqref{DT-splitting} in Theorem~\ref{corol-modified-DT-MAC} are inspired by the joint dependence-testing and splitting dependence-testing (DT) bounds for the discrete MAC, respectively~\cite{ML-ISIT12}. The latter is a loosening of the former that results from \emph{splitting} the joint outage event via a union bound. The joint-outage result~\eqref{DT-joint} provides a tighter bound on the ultimate performance, and the splitting-outage result~\eqref{DT-splitting} enables simpler evaluation. These two forms will be used in proving the second-order regions presented in Theorems~\ref{GMAC-joint} and~\ref{GMAC-splitting}, respectively.

\begin{proof} 
The two channel encoders randomly generate $M_1$ and $M_2$~codewords independently according to some given $n$-letter distributions $P_{X_1^n}$ and~$P_{X_2^n}$, respectively, where $n$ is the designated blocklength. Observing the output~$y^n$, the decoder chooses the first codeword pair $x_1^n(\hat{m}_1)$ and~$x_2^n(\hat{m}_2)$ that look ``jointly typical'' with $y^n$ in a \emph{modified} one-sided sense, specifically  satisfying all three of the following conditions:
\begin{subequations}
\begin{align}
\tilde{i}(x_1^n(\hat{m}_1);y^n|x_2^n(\hat{m}_2))>\log \gamma_1(x_1^n(\hat{m}_1),x_2^n(\hat{m}_2)), \\
\tilde{i}(x_2^n(\hat{m}_2);y^n|x_1^n(\hat{m}_1))>\log \gamma_2(x_1^n(\hat{m}_1),x_2^n(\hat{m}_2)), \\
\tilde{i}(x_1^n(\hat{m}_1),x_2^n(\hat{m}_2);y^n)>\log \gamma_3(x_1^n(\hat{m}_1),x_2^n(\hat{m}_2)),
\end{align}
\label{mod-decod-MAC}%
\end{subequations}
where $\gamma_1(x_1^n,x_2^n)$, $\gamma_2(x_1^n,x_2^n)$, $\gamma_3(x_1^n,x_2^n)$ are codeword-dependent thresholds and $\tilde{i}(x_1^n;y^n|x_2^n)$, $\tilde{i}(x_2^n;y^n|x_1^n)$, $\tilde{i}(x_1^n,x_2^n;y^n)$ are the corresponding realizations of the \emph{modified} mutual information random variables~$\tilde{i}(X_1^n;Y^n|X_2^n)$, $\tilde{i}(X_2^n;Y^n|X_1^n)$, $\tilde{i}(X_1^nX_2^n;Y^n)$, respectively.
The error probability averaged over the set of $M_1$ codewords of all possible realizations of the first codebook and the set of $M_2$ codewords of all possible realizations of the second codebook can then be bounded, similar to~\eqref{typ-1}-\eqref{typ-2}, as the sum of a joint-outage probability and three confusion probabilities as follows:
\allowdisplaybreaks{
\begin{align}
&\epsilon\!\leq\!P_{X_1^n}P_{X_2^n}P_{Y^n|X_1^nX_2^n}\left[\,\tilde{i}(X_1^n;Y^n|X_2^n)\leq \log\gamma_1(X_1^n,X_2^n) \right. \notag \\
& \qquad \qquad \qquad \qquad \qquad \cup\left. \tilde{i}(X_2^n;Y^n|X_1^n)\leq \log\gamma_2(X_1^n,X_2^n) \right. \notag \\
& \qquad \qquad \qquad \qquad \qquad \cup\left. \tilde{i}(X_1^nX_2^n;Y^n)\leq \log\gamma_3(X_1^n,X_2^n)\right] \notag\\
&+\frac{M_1-1}{2}P_{X_1^n}P_{X_2^n}P_{Y^n|X_2^n}\left[\,\tilde{i}(X_1^n;Y^n|X_2^n)>\log\gamma_1(X_1^n,X_2^n)\right] \notag\\
&+\frac{M_2-1}{2}P_{X_1^n}P_{X_2^n}P_{Y^n|X_1^n}\left[\,\tilde{i}(X_2^n;Y^n|X_1^n)>\log\gamma_2(X_1^n,X_2^n)\right] \notag\\
&+\frac{(M_1-1)(M_2-1)}{2}P_{X_1^n}P_{X_2^n}P_{Y^n}\left[\,\tilde{i}(X_1^nX_2^n;Y^n)>\log\gamma_3(X_1^n,X_2^n)\right].
\end{align}}
Applying the change of measure technique of~\eqref{change-measure-MAC} with the definitions~\eqref{unif-bound-MAC} yields
\begin{align}
\epsilon \leq\; &P_{X_1^n}P_{X_2^n}P_{Y^n|X_1^nX_2^n}\left[\,\tilde{i}(X_1^n;Y^n|X_2^n)\leq \log\gamma_1(X_1^n,X_2^n) \right. \notag \\
& \qquad \qquad \qquad \qquad \quad \cup\left. \tilde{i}(X_2^n;Y^n|X_1^n)\leq \log\gamma_2(X_1^n,X_2^n) \right. \notag \\
& \qquad \qquad \qquad \qquad \quad \cup\left. \tilde{i}(X_1^nX_2^n;Y^n)\leq \log\gamma_3(X_1^n,X_2^n)\right] \notag\\
&+K_{1n}\frac{M_1-1}{2}P_{X_1^n}P_{X_2^n}Q^{(1)}_{Y^n|X_2^n}\left[\,\tilde{i}(X_1^n;Y^n|X_2^n)>\log\gamma_1(X_1^n,X_2^n)\right] \notag\\
&+K_{2n}\frac{M_2-1}{2}P_{X_1^n}P_{X_2^n}Q^{(2)}_{Y^n|X_1^n}\left[\,\tilde{i}(X_2^n;Y^n|X_1^n)>\log\gamma_2(X_1^n,X_2^n)\right] \notag\\
&+K_{3n}\frac{(M_1-1)(M_2-1)}{2}P_{X_1^n}P_{X_2^n}Q^{(3)}_{Y^n}\left[\,\tilde{i}(X_1^nX_2^n;Y^n)>\log\gamma_3(X_1^n,X_2^n)\right].
\end{align}
The input cost constraints can be handled, analogously to the P2P case, by selecting all the decoding thresholds to be infinite~$\gamma_1(x_1^n,x_2^n)=\infty$, $\gamma_2(x_1^n,x_2^n)=\infty$ and $\gamma_3(x_1^n,x_2^n)=\infty$ if either $x_1^n\notin\mathcal{F}_{1n}$ or $x_2^n\notin\mathcal{F}_{2n}$, and selecting $\gamma_1(x_1^n,x_2^n)=\gamma_{1n}$, $\gamma_2(x_1^n,x_2^n)=\gamma_{2n}$ and $\gamma_3(x_1^n,x_2^n)=\gamma_{3n}$, otherwise. To handle the input cost constraints, analogously to~\eqref{DT-cost-1}-\eqref{DT-cost-PPV}, we obtain 
\begin{align}
\epsilon \leq\; &P_{X_1^n}P_{X_2^n}\left[X_1^n\notin\mathcal{F}_{1n} \cup X_2^n\notin\mathcal{F}_{2n}\right] \notag \\
&+ P_{X_1^n}P_{X_2^n}P_{Y^n|X_1^nX_2^n}\left[\,\tilde{i}(X_1^n;Y^n|X_2^n)\leq \log\gamma_{1n} \right. \notag \\
& \qquad \qquad \qquad \qquad \qquad \cup\left. \tilde{i}(X_2^n;Y^n|X_1^n)\leq \log\gamma_{2n} \right. \notag \\
& \qquad \qquad \qquad \qquad \qquad \cup\left. \tilde{i}(X_1^nX_2^n;Y^n)\leq \log\gamma_{3n}\;\right] \notag\\
&+K_{1n}\frac{M_1-1}{2}P_{X_1^n}P_{X_2^n}Q^{(1)}_{Y^n|X_2^n}\left[\,\tilde{i}(X_1^n;Y^n|X_2^n)>\log\gamma_{1n}\right] \notag\\
&+K_{2n}\frac{M_2-1}{2}P_{X_1^n}P_{X_2^n}Q^{(2)}_{Y^n|X_1^n}\left[\,\tilde{i}(X_2^n;Y^n|X_1^n)>\log\gamma_{2n}\right] \notag\\
&+K_{3n}\frac{(M_1-1)(M_2-1)}{2}P_{X_1^n}P_{X_2^n}Q^{(3)}_{Y^n}\left[\,\tilde{i}(X_1^nX_2^n;Y^n)>\log\gamma_{3n}\right].
\label{joint DT for MAC}
\end{align}
To simplify the analysis, one could apply a union bound to the second term on the RHS of~\eqref{joint DT for MAC}, the joint outage event, to obtain the following potentially looser, but simpler, outage-splitting bound. 
\begin{align}
\epsilon \leq\; & P_{X_1^n}P_{X_2^n}\left[X_1^n\notin\mathcal{F}_{1n} \cup X_2^n\notin\mathcal{F}_{2n}\right] \notag \\
&+P_{X_1^n}P_{X_2^n}P_{Y^n|X_1^nX_2^n}\left[\,\tilde{i}(X_1^n;Y^n|X_2^n)\leq \log\gamma_{1n}\right] \notag\\
&+K_{1n}\frac{M_1-1}{2}P_{X_1^n}P_{X_2^n}Q^{(1)}_{Y^n|X_2^n}\left[\,\tilde{i}(X_1^n;Y^n|X_2^n)>\log\gamma_{1n}\right] \notag\\
&+ P_{X_1^n}P_{X_2^n}P_{Y^n|X_1^nX_2^n}\left[\,\tilde{i}(X_2^n;Y^n|X_1^n)\leq \log\gamma_{2n}\right] \notag\\
&+K_{2n}\frac{M_2-1}{2}P_{X_1^n}P_{X_2^n}Q^{(2)}_{Y^n|X_1^n}\left[\,\tilde{i}(X_2^n;Y^n|X_1^n)>\log\gamma_{2n}\right] \notag\\
&+ P_{X_1^n}P_{X_2^n}P_{Y^n|X_1^nX_2^n}\left[\,\tilde{i}(X_1^nX_2^n;Y^n)\leq \log\gamma_{3n}\right] \notag\\
&+K_{3n}\frac{(M_1-1)(M_2-1)}{2}P_{X_1^n}P_{X_2^n}Q^{(3)}_{Y^n}\left[\,\tilde{i}(X_1^nX_2^n;Y^n)>\log\gamma_{3n}\right]. 
\label{Splitting DT for MAC}
\end{align}
Upon remapping all of the non-feasible codewords of each codebook to arbitrary sequences~$x_1^n(0)\in\mathcal{F}_{1n}$ or $x_2^n(0)\in\mathcal{F}_{2n}$, respectively, without modifying the decoding regions, we infer that there exists a pair of deterministic codebooks with $M_1$ codewords belonging to the feasible set~$\mathcal{F}_{1n}$ and $M_2$ codewords belonging to the feasible set~$\mathcal{F}_{2n}$, respectively, whose average error probability~$\epsilon$ satisfies~\eqref{joint DT for MAC} or~\eqref{Splitting DT for MAC}.

To conclude the final assertion~\eqref{th-MAC} of Theorem~\ref{corol-modified-DT-MAC}, it is sufficient to observe that the last six summands on the RHS of \eqref{Splitting DT for MAC} are three weighted sums of two types of error in three Bayesian binary hypothesis tests, respectively, and therefore correspond to the average error probabilities of these tests. The optimal test for each case is an LRT, as we have seen in~\eqref{mod-decod-MAC}, with the optimal threshold equal to the ratio of priors or simply the ratio of the coefficients of the two error probabilities of the test, as given in~\eqref{th-MAC}. 
\end{proof}

\subsection{Second-Order Characterization for the Gaussian MAC}
\label{sub-2nd-mac}

In this section, we specialize the achievability bound of Theorem 4 to the Gaussian MAC and prove Theorems~2 and 3. Our approach is analogous to that taken for the P2P Gaussian channel.

\subsubsection{Coding on Independent Power Shells}

First, we choose the pair of input distributions to be independent uniform distributions on the respective power shells
\begin{align}
P_{X_1^nX_2^n}(x_1^n,x_2^n)&=\frac{\delta(||x_1^n||-\sqrt{nP_1})}{S_n(\sqrt{nP_1})}\cdot\frac{\delta(||x_2^n||-\sqrt{nP_2})}{S_n(\sqrt{nP_2})}, \label{unif-shell-mac}
\end{align}
with the same notations as in~\eqref{unif-shell}. Note that this pair of distributions satisfies the input power constraint with probability one, that is,
\begin{align}
P_{X_1^n}P_{X_2^n}\left[X_1^n\notin\mathcal{F}_{1n} \cup X_2^n\notin\mathcal{F}_{2n}\right]=P_{X_1^n}P_{X_2^n}\left[||X_1^n||^2>nP_1 \cup ||X_2^n||^2>nP_2\right]=0.
\label{cost=0-MAC}
\end{align}
Moreover, analogous to~\eqref{shell-output} for the P2P Gaussian channel, the conditional output distributions induced by this input pair are
\begin{align}
P_{Y^n|X_2^n}(y^n|x_2^n) &= \frac{1}{2}\pi^{-n/2}\Gamma\left(\frac{n}{2}\right) e^{-nP_1/2} e^{-||y^n-x_2^n||^2/2}  \frac{I_{n/2-1}(||y^n-x_2^n||\sqrt{nP_1})}{(||y^n-x_2^n||\sqrt{nP_1})^{n/2-1}},  \label{shell-output-1} \\
P_{Y^n|X_1^n}(y^n|x_1^n) &= \frac{1}{2}\pi^{-n/2}\Gamma\left(\frac{n}{2}\right) e^{-nP_2/2} e^{-||y^n-x_1^n||^2/2}   \frac{I_{n/2-1}(||y^n-x_1^n||\sqrt{nP_2})}{(||y^n-x_1^n||\sqrt{nP_2})^{n/2-1}},  \label{shell-output-2} 
\end{align}
where $I_v(\cdot)$ is again the modified Bessel function of the first kind and~$v$-th order. 

The analysis of the unconditional output distribution~$P_{Y^n}$ for such an input pair is more complicated, and appears unlikely to be expressed in closed form.\footnote{In fact, our former expression~\cite[Eq. (39)]{ML-Allerton12} for this induced output distribution appears to be incorrect, and we have not been able to obtain a closed form expression for this distribution, even using Bessel functions and the like.} However, we can fully characterize the distribution $U^n:=X_1^n+X_2^n$ of  the superimposed input to the channel, and use this distribution for our later analysis. In particular, under the independent uniform distribution~\eqref{unif-shell-mac} for $X_1^n$ and $X_2^n$ on the respective power shells, we have $P_{U^n}(u^n)=0$ for any $u^n\in\mathbb{R}^n$ that satisfies $||u^n||<|\sqrt{nP_1}-\sqrt{nP_2}|$ or $||u^n||>\sqrt{nP_1}+\sqrt{nP_2}$. Moreover, we have $P_{U^n}(u^n)=0$ for those $u^n\in\mathbb{R}^n$ satisfying $||u^n||=|\sqrt{nP_1}-\sqrt{nP_2}|$, since
\begin{align}
\Pr\left[||U^n||<\left|\sqrt{nP_1}-\sqrt{nP_2}\right|\right]&=\Pr\left[||X_1^n+X_2^n||<\left|\sqrt{nP_1}-\sqrt{nP_2}\right|\right]=0,
\end{align}
and
\begin{align}
\Pr\left[||U^n||\leq\left|\sqrt{nP_1}-\sqrt{nP_2}\right|\right] &= \Pr\left[||U^n||=\left|\sqrt{nP_1}-\sqrt{nP_2}\right|\right] \\
&=\Pr\left[||X_1^n+X_2^n||=\left|\sqrt{nP_1}-\sqrt{nP_2}\right|\right] \\
&=\mathbb{E}_{X_2^n}\left[\Pr\left[||X_1^n+x_2^n||=\left|\sqrt{nP_1}-\sqrt{nP_2}\right|\right]\right] \\
&=\mathbb{E}_{X_2^n}\left[\Pr\left[X_1^n=-\sqrt{nP_1}\frac{x_2^n}{||x_2^n||}\right]\right] \\
&=\mathbb{E}_{X_2^n}[0] =0.
\end{align}
Analogously, we have $P_{U^n}(u^n)=0$ for those $u^n\in\mathbb{R}^n$ satisfying $||u^n||=\sqrt{nP_1}+\sqrt{nP_2}$, since
\begin{align}
\Pr\left[||U^n||\leq\sqrt{nP_1}+\sqrt{nP_2}\right]=1 \quad \text{and} \quad  \Pr\left[||U^n||<\sqrt{nP_1}+\sqrt{nP_2}\right] = 1.
\end{align}
However, for any $u^n\in\mathbb{R}^n$ belonging to the hollow sphere $|\sqrt{nP_1}-\sqrt{nP_2}|<||u^n||<\sqrt{nP_1}+\sqrt{nP_2}$, we have
\begin{align}
P_{U^n}(u^n) & =\int_{\mathbb{R}^n}  P_{X_1^n}(x_1^n) P_{X_2^n}(u^n-x_1^n)  dx_1^n  \\
& = \int_{\mathbb{R}^n} \frac{\delta(||x_1^n||-\sqrt{nP_1})}{S_n(\sqrt{nP_1})} \cdot \frac{\delta(||u^n-x_1^n||-\sqrt{nP_2})}{S_n(\sqrt{nP_2})}  dx_1^n   \\
& =  \int_0^\pi \int_0^\infty  \frac{\delta(r-\sqrt{nP_1})}{S_n(\sqrt{nP_1})} \cdot \frac{\delta(\sqrt{||u^n||^2-2||u^n||r\cos\theta+r^2}-\sqrt{nP_2})}{S_n(\sqrt{nP_2})}   S_{n-1}(r\sin\theta) r dr d\theta  \label{decomp-mac} \\
& = \frac{\sqrt{nP_1}S_{n-1}(\sqrt{nP_1})}{S_n(\sqrt{nP_1})}  \int_0^\pi \frac{\delta\left(\sqrt{||u^n||^2-2||u^n||\sqrt{nP_1}\cos\theta+nP_1}-\sqrt{nP_2}\right)}{S_n(\sqrt{nP_2})} \left(\sin\theta\right)^{n-2}  d\theta dr \label{th0} \\
& =  \frac{1}{\sqrt{\pi}} \frac{\Gamma\left(\frac{n}{2}\right)}{\Gamma\left(\frac{n-1}{2}\right)} \int_0^\pi \frac{1}{S_n(\sqrt{nP_2})} \sqrt{\frac{P_2}{P_1}} \frac{\delta(\theta-\theta_0)}{||u^n||\sin\theta_0} \left(\sin\theta\right)^{n-2} dr \\
& =  \sqrt{\frac{P_2}{\pi P_1}} \frac{\Gamma\left(\frac{n}{2}\right)}{\Gamma\left(\frac{n-1}{2}\right)} \frac{\Gamma\left(\frac{n}{2}\right)}{2\pi^{n/2}(nP_2)^{(n-1)/2}}  \frac{1}{||u^n||} \left(1-\left(\frac{||u^n||^2+n(P_1-P_2)}{2\sqrt{nP_1}||u^n||}\right)^2\right)^{(n-3)/2},
\end{align}
where~\eqref{decomp-mac} follows from a decomposition of the space~$\mathbb{R}^n$ into a continuum of ring elements as in~\eqref{decomp}, and~\eqref{th0} follows from the identity $\delta(g(x))=\frac{\delta(x-x_0)}{|g'(x_0)|}$ with $x_0$ being the real root of $g(x)$, so that
\begin{align}
\delta\left(\sqrt{||u^n||^2-2||u^n||\sqrt{nP_1}\cos\theta+nP_1}-\sqrt{nP_2}\right) &=  \frac{\delta(\theta-\theta_0)}{\left|\frac{2||u^n||\sqrt{nP_1}\sin\theta_0}{2\sqrt{nP_2}}\right|} = \sqrt{\frac{P_2}{P_1}} \frac{\delta(\theta-\theta_0)}{||u^n||\sin\theta_0},
\end{align}
in which $\theta_0\in(0,\pi)$ is defined as the solution to 
\begin{align}
\cos\theta_0= \frac{||u^n||^2+n(P_1-P_2)}{2\sqrt{nP_1}||u^n||}.
\end{align}
The unconditional output distribution $P_{Y^n}$ is now given by
\begin{align}
P_{Y^n}(y^n) =\int_{\mathbb{R}^n}  P_{U^n}(u^n) P_{Y^n|U^n}(y^n|u^n)  du^n, \label{shell-output-3} 
\end{align}
where $P_{Y^n|U^n}(y^n|u^n)$ is the i.i.d. Gaussian distribution~$\mathcal{N}(y^n; u^n, I_n)$ of the channel noise.

Next, we choose the triple of (conditional) output distributions to be the capacity-achieving output distributions with respect to each case, that is,
\begin{align}
Q^{(1)}_{Y^n|X_2^n}(y^n|x_2^n)&\sim\mathcal{N}(y^n; x_2^n,(1+P_1)I_n), \label{Q1}\\
Q^{(2)}_{Y^n|X_1^n}(y^n|x_1^n)&\sim\mathcal{N}(y^n; x_1^n,(1+P_2)I_n), \label{Q2}\\
Q^{(3)}_{Y^n}(y^n)&\sim\mathcal{N}(y^n; \mathbf{0},(1+P_1+P_2)I_n). \label{Q3}
\end{align}
The following proposition will then bound the R-N derivatives introduced in~\eqref{unif-bound-MAC}. The proof, which is a slight generalization of the one for the P2P case, is given in Appendix~\ref{proof-Prop-MAC}.

\begin{proposition}
\label{divergence-bound-MAC}
Let $P_{Y^n|X_2^n},P_{Y^n|X_1^n},P_{Y^n}$ be the (conditional) distributions~\eqref{shell-output-1}, \eqref{shell-output-2}, \eqref{shell-output-3} induced on the output of the Gaussian MAC by a pair of independent uniform input distributions on the respective power shells~\eqref{unif-shell-mac}, and let $Q^{(1)}_{Y^n|X_2^n},Q^{(2)}_{Y^n|X_1^n},Q^{(3)}_{Y^n}$ be the (conditional) capacity-achieving output distributions~\eqref{Q1}, \eqref{Q2}, \eqref{Q3}. There exist positive constants~$K_1,K_2,K_3$ such that, for any $x_1^n,x_2^n,y^n\in\mathbb{R}^n$, and for sufficiently large~$n$,
\begin{subequations}
\begin{align}
\frac{dP_{Y^n|X_2^n}(y^n|x_2^n)}{dQ^{(1)}_{Y^n|X_2^n}(y^n|x_2^n)}&\leq K_1, \label{unif-mac-1}\\
\frac{dP_{Y^n|X_1^n}(y^n|x_1^n)}{dQ^{(2)}_{Y^n|X_1^n}(y^n|x_1^n)}&\leq K_2, \label{unif-mac-2} \\
\frac{dP_{Y^n}(y^n)}{dQ^{(3)}_{Y^n}(y^n)} & \leq K_3. \label{unif-mac-3}
\end{align}
\end{subequations}
\end{proposition}
\textit{Remark.} Using some more complicated manipulations, the proposition can be shown to be valid for any finite~$n$, but the above statement is enough for our second-order analysis.

Proposition~\ref{divergence-bound-MAC} facilitates the use of Theorem~\ref{corol-modified-DT-MAC} with the aforementioned choices for the input distributions and the reference output distributions. Substituting~\eqref{cost=0-MAC} into the achievability bounds~\eqref{DT-joint} and~\eqref{DT-splitting} of Theorem~\ref{corol-modified-DT-MAC} leaves only the confusion and joint/individual outage probabilities. Note that, for simplicity of analysis, we will use the choice of thresholds as indicated in~\eqref{th-MAC} for both of the joint-outage and outage-splitting bounds above, although it need not be the optimal choice for the joint case. 
In the following, we evaluate the outage and confusion probabilities for sufficiently large blocklength to obtain the second-order achievable bounds.

\subsubsection{Evaluation of the Outage Probability}
\label{sub-MAC-outage}

The joint outage probability for the Gaussian MAC can be written in the following generic form
\begin{align}
P_{X_1^n}P_{X_2^n}P_{Y^n|X_1^nX_2^n}&\left[\,\tilde{i}(X_1^n;Y^n|X_2^n)\leq \log\gamma_1 \right. \notag \\
&  \quad \cup\left. \tilde{i}(X_2^n;Y^n|X_1^n)\leq \log\gamma_2 \right. \notag \\
&  \quad \cup\left. \tilde{i}(X_1^nX_2^n;Y^n)\leq \log\gamma_3\;\right] \\
&=1 - P_{X_1^n}P_{X_2^n}P_{Y^n|X_1^nX_2^n}\left[\,\tilde{\mathbf{i}}(X_1^nX_2^n;Y^n) > 
\begin{pmatrix}
\log\gamma_1 \\
\log\gamma_2 \\
\log\gamma_3
\end{pmatrix}
\right] ,
\label{outage-pr-joint}
\end{align}
in which the modified mutual information random \textit{vector} is defined as
\begin{align}
\tilde{\mathbf{i}}(X_1^nX_2^n;Y^n):=
\begin{pmatrix}
\tilde{i}(X_1^n;Y^n|X_2^n) \\
\tilde{i}(X_2^n;Y^n|X_1^n) \\
\tilde{i}(X_1^nX_2^n;Y^n)
\end{pmatrix}, \label{mi-mac}
\end{align}
and the input distribution~$P_{X_1^n}P_{X_2^n}$ used in the outage formulation above is the independent uniform distribution on the respective power shells~\eqref{unif-shell-mac}, and the vector inequality in~\eqref{outage-pr-joint} is understood as being element wise.

Under the $P_{Y^n|X_1^nX_2^n}$ channel law, the output~$Y^n$ can be written in the form
\begin{align}
Y^n=X_1^n+X_2^n+Z^n,
\end{align}
where $Z^n\sim \mathcal{N}(\mathbf{0},I_n)$ is the i.i.d. unit-variance channel noise. With the choices~\eqref{Q1} and \eqref{Q2} for $Q^{(1)}_{Y^n|X_2^n}$ and $Q^{(2)}_{Y^n|X_1^n}$, respectively, the first two elements of this random vector simplify analogous to~\eqref{mi-p2p} as follows:
\begin{align}
\tilde{i}(X_1^n;Y^n|X_2^n)&\equiv nC(P_1)+\frac{\log e}{2(1+P_1)}\left[P_1(n-||Z^n||^2)+2 \langle X_1^n,Z^n \rangle\right], \label{mi-mac-1}\\
\tilde{i}(X_2^n;Y^n|X_1^n)&\equiv nC(P_2)+\frac{\log e}{2(1+P_2)}\left[P_2(n-||Z^n||^2)+2 \langle X_2^n,Z^n \rangle\right]. \label{mi-mac-2}
\end{align}
Moreover, with the choice~\eqref{Q3} for $Q^{(3)}_{Y^n}$, the third element of the modified mutual information random vector also simplifies to
\begin{align}
\tilde{i}(X_1^n,X_2^n;Y^n)&\equiv \log\frac{(2\pi)^{-n/2}e^{-||Y^n-X_1^n-X_2^n||^2/2}}{(2\pi(1+P_1+P_2))^{-n/2}e^{-||Y^n||^2/2(1+P_1+P_2)}} \\
&=\frac{n}{2}\log(1+P_1+P_2)+\frac{\log e}{2}\left[\frac{||Y^n||^2}{1+P_1+P_2}-||Y^n-X_1^n-X_2^n||^2\right] \\
&=nC(P_1+P_2)+\frac{\log e}{2(1+P_1+P_2)}\left[||X_1^n+X_2^n+Z^n||^2-(1+P_1+P_2)||Z^n||^2\right] \\
&=nC(P_1+P_2) +\frac{\log e}{2(1+P_1+P_2)}\left[(P_1+P_2)(n-||Z^n||^2) + 2 \langle X_1^n, X_2^n \rangle+2 \langle X_1^n,Z^n \rangle+2 \langle X_2^n,Z^n \rangle\right], \label{mi-mac-3}
\end{align}
since $||X_1^n||^2=nP_1$ and $||X_2^n||^2=nP_2$ with probability one.

Note that, although these random variables are written in the form of summations, the summands are not independent, since neither of the inputs $X_1^n$ and $X_2^n$ are independent across time. Therefore, a direct application of the conventional CLT is not possible. Moreover, the symmetry arguments used in the Gaussian P2P case \cite{PPV,Hayashi} do not apply, since the realization of the inner product RV~$\langle X_1^n, X_2^n \rangle$ varies with different pairs of codewords~$(x_1^n,x_2^n)$ on the power shells.

However, recall that independent uniform RVs on the power shells can be viewed as functions of i.i.d. Gaussian RVs. More precisely, let $W_1^n\sim\mathcal{N}(\mathbf{0},I_n)$ and $W_2^n\sim \mathcal{N}(\mathbf{0},I_n)$ be i.i.d. Gaussian RVs independent of each other and the channel noise~$Z^n\sim \mathcal{N}(\mathbf{0},I_n)$. The elements $X_{1t}$ and $X_{2t}$, $t=1,...,n$, of the independent uniformly distributed RVs $X_1^n,X_2^n$ on the power shells~\eqref{unif-shell-mac} can be expressed as
\begin{align}
X_{1t}=\sqrt{nP_1}\frac{W_{1t}}{||W_1^n||}, \qquad 
X_{2t}=\sqrt{nP_2}\frac{W_{2t}}{||W_2^n||}.
\end{align}
Hence, we can apply the CLT for functions of Proposition~\ref{CLT-func} as follows. Consider the vector $\{\mathbf{U}_t=(U_{1t},...,U_{6t})\}_{t=1}^\infty$ whose elements are
\begin{align}
U_{1t}&=1-Z_t^2, \\
U_{2t}&=\sqrt{P_1}W_{1t}Z_{t}, \\
U_{3t}&=\sqrt{P_2}W_{2t}Z_{t}, \\
U_{4t}&=\sqrt{P_1P_2}W_{1t}W_{2t}, \\
U_{5t}&=W_{1t}^2-1, \\
U_{6t}&=W_{2t}^2-1. 
\end{align}
Note that this random vector has an i.i.d. distribution across time $t=1,...,n$, and its moments can be easily verified to satisfy $\mathbb{E}[\mathbf{U}_1]=0$ and $\mathbb{E}[||\mathbf{U}_1||_2^3]<\infty$. Moreover, the covariance matrix of this vector is given by
\begin{align}
\text{Cov}(\mathbf{U}_1)=
\begin{pmatrix}
2 & 0 & 0 & 0 & 0 & 0 \\
0 & P_1 & 0 & 0 & 0 & 0 \\
0 & 0 & P_2 & 0 & 0 & 0 \\
0 & 0 & 0 & P_1P_2 & 0 & 0 \\
0 & 0 & 0 & 0 & 2 & 0 \\
0 & 0 & 0 & 0 & 0 & 2 
\end{pmatrix}.
\end{align}

Next, define the vector function $\mathbf{f}(\mathbf{u})=(f_1(\mathbf{u}),f_2(\mathbf{u}),f_3(\mathbf{u}))$ whose three components are
\allowdisplaybreaks{
\begin{align}
f_1(\mathbf{u})&=P_1u_1+ \frac{2u_2}{\sqrt{1+u_5}}, \\
f_2(\mathbf{u})&=P_2u_1+ \frac{2u_3}{\sqrt{1+u_6}}, \\
f_3(\mathbf{u})&=(P_1+P_2)u_1+ \frac{2u_2}{\sqrt{1+u_5}} + \frac{2u_3}{\sqrt{1+u_6}} + \frac{2u_4}{\sqrt{1+u_5}\sqrt{1+u_6}}.
\end{align}}%
Again, $\mathbf{f}(\mathbf{0})= \mathbf{0}$ and all the first- and second-order partial derivatives of all three components of $\mathbf{f}$ are continuous in a neighborhood of $\mathbf{u}=\mathbf{0}$. Moreover, the Jacobian matrix $\{\frac{\partial f_l(\mathbf{u})}{\partial u_j}\}_{3\times6}$ at $\mathbf{u}=\mathbf{0}$ can be readily verified to be
\begin{align}
\left.J\right|_{\mathbf{u}=\mathbf{0}}=
\begin{pmatrix}
P_1 & 2 & 0 & 0 & 0 & 0 \\
P_2 & 0 & 2 & 0 & 0 & 0 \\
P_1+P_2 & 2 & 2 & 2 & 0 & 0
\end{pmatrix}.
\end{align}
Moreover, the first two components, similar to the P2P case~\eqref{f-sum-1}-\eqref{f-sum-3}, give 
\begin{align}
f_1\left(\frac{1}{n}\sum_{t=1}^n \mathbf{U}_t\right) &= \frac{1}{n}\left[P_1(n-||Z^n||^2)+2 \langle X_1^n,Z^n \rangle\right], \\
f_2\left(\frac{1}{n}\sum_{t=1}^n \mathbf{U}_t\right) &= \frac{1}{n}\left[P_2(n-||Z^n||^2)+2 \langle X_2^n,Z^n \rangle\right],
\end{align}
and the third component yields 
\begin{align}
f_3\left(\frac{1}{n}\sum_{t=1}^n \mathbf{U}_t\right)
&=\frac{P_1+P_2}{n}\sum_{t=1}^n (1-Z_t^2)+ \frac{2\frac{1}{n}\sum_{t=1}^n \sqrt{P_1}W_{1t}Z_{t}}{\sqrt{1+\frac{1}{n}\sum_{t=1}^n (W_{1t}^2-1)}} 
+ \frac{2\frac{1}{n}\sum_{t=1}^n \sqrt{P_2}W_{2t}Z_{t}}{\sqrt{1+\frac{1}{n}\sum_{t=1}^n (W_{2t}^2-1)}} \notag \\
& \qquad \qquad + \frac{2\frac{1}{n}\sum_{t=1}^n \sqrt{P_1P_2}W_{1t}W_{2t}}{\sqrt{1+\frac{1}{n}\sum_{t=1}^n (W_{1t}^2-1)}\sqrt{1+\frac{1}{n}\sum_{t=1}^n (W_{2t}^2-1)}} \\
&= \frac{1}{n}\sum_{t=1}^n (P_1+P_2)(1-Z_t^2)+ \frac{2}{n}\sum_{t=1}^n  \frac{\sqrt{nP_1}W_{1t}}{||W_1^n||}Z_{t} + \frac{2}{n}\sum_{t=1}^n  \frac{\sqrt{nP_2}W_{2t}}{||W_2^n||}Z_{t}  \notag \\
& \qquad \qquad + \frac{2}{n}\sum_{t=1}^n  \frac{\sqrt{nP_1}W_{1t}}{||W_1^n||^2} \frac{\sqrt{nP_2}W_{2t}}{||W_2^n||^2} \\
&= \frac{1}{n}\left[(P_1+P_2)(n-||Z^n||^2) + 2 \langle X_1^n, X_2^n \rangle +2 \langle X_1^n,Z^n \rangle +2 \langle X_2^n,Z^n \rangle\right].
\end{align}

Recalling~\eqref{mi-mac-1}, \eqref{mi-mac-2}, \eqref{mi-mac-3}, we now conclude from Proposition~\ref{CLT-func} that the modified mutual information random vector~\eqref{mi-mac} converges in distribution to a 3-dimensional Gaussian random vector with mean vector $n\mathbf{C}(P_1,P_2)$ and covariance matrix given by  
\begin{align}
&\frac{1}{n}\left(\frac{n\log e}{2}\right)^2 
\begin{pmatrix}
\frac{1}{1+P_1}&\!\!0 &\!\! 0 \\
0&\!\!\frac{1}{1+P_2} &\!\! 0 \\
0&\!\!0 & \!\!\frac{1}{1+P_1+P_2}
\end{pmatrix} \times \\
&\qquad \begin{pmatrix}
P_1 & \!\!2 & 0 & 0 & 0 & 0 \\
P_2 & \!\!0 & 2 & 0 & 0 & 0 \\
P_1+P_2 & \!\!2 & 2 & 2 & 0 & 0
\end{pmatrix}
\begin{pmatrix}
2 & 0 & 0 & 0 & 0 & 0 \\
0 & P_1 & 0 & 0 & 0 & 0 \\
0 & 0 & P_2 & 0 & 0 & 0 \\
0 & 0 & 0 & P_1P_2 & 0 & 0 \\
0 & 0 & 0 & 0 & 2 & 0 \\
0 & 0 & 0 & 0 & 0 & 2 
\end{pmatrix}
\begin{pmatrix}
P_1 & P_2 & \!\!P_1+P_2 \\ 
2 & 0 & \!\!2  \\
0 & 2 & \!\!2 \\
0 & 0 & \!\!2 \\
0 & 0 & \!\!0 \\
0 & 0 & \!\!0
\end{pmatrix} 
\begin{pmatrix}
\frac{1}{1+P_1} & \!\!0 & \!\!0 \\
0 & \!\!\frac{1}{1+P_2} & \!\!0 \\
0 & \!\!0 & \!\!\frac{1}{1+P_1+P_2}
\end{pmatrix} \\
&=\frac{n\log^2 e}{2}
\begin{pmatrix}
\frac{P_1(P_1+2)}{(P_1+1)^2} & \frac{P_1P_2}{(P_1+1)(P_2+1)} & \frac{P_1(P_1+P_2+2)}{(P_1+1)(P_1+P_2+1)}  \\
\frac{P_1P_2}{(P_1+1)(P_2+1)} & \frac{P_2(P_2+2)}{(P_2+1)^2} & \frac{P_2(P_1+P_2+2)}{(P_2+1)(P_1+P_2+1)}  \\
\frac{P_1(P_1+P_2+2)}{(P_1+1)(P_1+P_2+1)} & \frac{P_2(P_1+P_2+2)}{(P_2+1)(P_1+P_2+1)} & \frac{(P_1+P_2)(P_1+P_2+2)+2P_1P_2}{(P_1+P_2+1)^2} 
\end{pmatrix} = n\mathbf{V}(P_1,P_2).
\end{align}
In particular, the joint outage probability is bounded as 
\begin{align}
P_{X_1^n}P_{X_2^n}P_{Y^n|X_1^nX_2^n}&\left[\,\tilde{i}(X_1^n;Y^n|X_2^n)\leq \log\left(K_1\frac{M_1-1}{2}\right) \right. \notag \\
&  \quad \cup\left. \tilde{i}(X_2^n;Y^n|X_1^n)\leq \log\left(K_2\frac{M_2-1}{2}\right) \right. \notag \\
&  \quad \cup\left. \tilde{i}(X_1^nX_2^n;Y^n)\leq \log\left(K_3\frac{(M_1-1)(M_2-1)}{2}\right)\right] \\
&\leq 1- \Pr\left[\mathcal{N}(n\mathbf{C}(P_1,P_2),n\mathbf{V}(P_1,P_2))>\begin{pmatrix}
\log\left(K_1\frac{M_1-1}{2}\right) \\
\log\left(K_2\frac{M_2-1}{2}\right) \\
\log\left(K_3\frac{(M_1-1)(M_2-1)}{2}\right)
\end{pmatrix}\right] + \frac{B_1}{\sqrt{n}}. 
\label{outage-mac-joint}%
\end{align}
where $B_1$ is the constant introduced in Proposition~\ref{CLT-func}.
Moreover, the individual outage probabilities are bounded as
\begin{subequations}
\begin{align}
&P_{X_1^n}P_{X_2^n}P_{Y^n|X_1^nX_2^n}\left[\,\tilde{i}(X_1^n;Y^n|X_2^n)\leq \log\left(K_1\frac{M_1-1}{2}\right) \right] \leq  Q\left(\frac{nC(P_1)-\log\left(K_1\frac{M_1-1}{2}\right)}{\sqrt{nV(P_1)}}\right) + \frac{B_{11}}{\sqrt{n}}, \label{outage-mac-1} \\
&P_{X_1^n}P_{X_2^n}P_{Y^n|X_1^nX_2^n}\left[\,\tilde{i}(X_2^n;Y^n|X_1^n)\leq \log\left(K_2\frac{M_2-1}{2}\right) \right] \leq  Q\left(\frac{nC(P_2)-\log\left(K_2\frac{M_2-1}{2}\right)}{\sqrt{nV(P_2)}}\right) + \frac{B_{12}}{\sqrt{n}}, \label{outage-mac-2} \\
&P_{X_1^n}P_{X_2^n}P_{Y^n|X_1^nX_2^n}\left[\,\tilde{i}(X_1^n;Y^n|X_2^n)\leq \log\left(K_3\frac{(M_1-1)(M_2-1)}{2}\right) \right]  \notag \\
&\qquad \qquad \qquad \qquad \qquad \qquad \qquad \qquad \qquad \quad \leq Q\left(\frac{nC(P_1+P_2)-\log\left(K_3\frac{(M_1-1)(M_2-1)}{2}\right)}{\sqrt{n[V(P_1+P_2)+V_3(P_1,P_2)]}}\right) + \frac{B_{13}}{\sqrt{n}}, \label{outage-mac-3}
\end{align} 
\label{outage-mac-splitting}%
\end{subequations}
where $B_{11}, B_{12}, B_{13}$ are also the constants introduced in Proposition~\ref{CLT-func}.

\subsubsection{Evaluation of the Confusion Probability}
\label{sub-2nd-Ach-MAC}

The confusion probabilities for the Gaussian MAC can be written in the following generic form 
\begin{subequations}
\begin{align}
&K_1\frac{M_1-1}{2}P_{X_1^n}P_{X_2^n}Q^{(1)}_{Y^n|X_2^n}\left[\,\tilde{i}(X_1^n;Y^n|X_2^n)> \log\left(K_1\frac{M_1-1}{2}\right)\right],  \\
&K_2\frac{M_2-1}{2}P_{X_1^n}P_{X_2^n}Q^{(2)}_{Y^n|X_1^n}\left[\,\tilde{i}(X_2^n;Y^n|X_1^n)>\log\left(K_2\frac{M_2-1}{2}\right)\right], \\
&K_3\frac{(M_1-1)(M_2-1)}{2}P_{X_1^n}P_{X_2^n}Q^{(3)}_{Y^n}\!\left[\,\tilde{i}(X_1^nX_2^n;Y^n)\!\!>\!\log\left(K_3\frac{(M_1-1)(M_2-1)}{2}\right)\right],
\end{align}
\end{subequations}
where $P_{X_1^n}P_{X_2^n}$ is the independent uniform input distribution on the respective power shells~\eqref{unif-shell-mac}, and $Q^{(1)}_{Y^n|X_2^n=x_2^n}$ $Q^{(2)}_{Y^n|X_1^n=x_1^n}$ and $Q^{(3)}_{Y^n}$ are the (conditional) capacity achieving output distributions~\eqref{Q1}, \eqref{Q2}, \eqref{Q3} for the Gaussian MAC. 

Focusing on the conditional confusion probabilities for fixed input sequences~$x_1^n$ and~$x_2^n$ on the respective power shells, we employ the change of measure technique of~\eqref{change of measure} with $P_{Y^n|X_1^n=x_1^n,X_2^n=x_2^n}$ in the role of~$P$, and $Q^{(1)}_{Y^n|X_2^n=x_2^n}$ $Q^{(2)}_{Y^n|X_1^n=x_1^n}$ and $Q^{(3)}_{Y^n}$ respectively in the role of~$Q$ to obtain the following refined large deviation bounds
\begin{align}
Q^{(1)}_{Y^n|X_2^n=x_2^n}\left[\,\tilde{i}(x_1^n;Y^n|x_2^n)> \log\gamma_1\right] &\leq \frac{B_{21}}{\sqrt{n}}\gamma_1^{-1}, \\
Q^{(2)}_{Y^n|X_1^n=x_1^n}\left[\,\tilde{i}(x_2^n;Y^n|x_1^n)> \log\gamma_2\right] &\leq \frac{B_{22}}{\sqrt{n}}\gamma_2^{-1},\\
Q^{(3)}_{Y^n}\!\left[\,\tilde{i}(x_1^n,x_2^n;Y^n)\!\!>\!\log\gamma_3\right] &\leq \frac{B_{23}}{\sqrt{n}}\gamma_3^{-1},
\end{align}
which follow from~\cite[Lemma 47]{PPV}. Specific expressions for the finite constants $B_{21},B_{22},B_{23}$ can be readily obtained in terms of the power constraints~$P_1$ and $P_2$, but are not  our main interest. Since these bounds are uniform with respect to the location of the input sequences~$x_1^n$ and~$x_2^n$ on the respective power shells, we can bound the unconditional confusion probabilities as
\begin{subequations}
\begin{align}
K_1\frac{M_1-1}{2}P_{X_1^n}P_{X_2^n}Q^{(1)}_{Y^n|X_2^n}\left[\,\tilde{i}(X_1^n;Y^n|X_2^n)> \log\left(K_1\frac{M_1-1}{2}\right)\right] &\leq \frac{B_{12}}{\sqrt{n}}, \\
K_2\frac{M_2-1}{2}P_{X_1^n}P_{X_2^n}Q^{(2)}_{Y^n|X_1^n}\left[\,\tilde{i}(X_2^n;Y^n|X_1^n)>\log\left(K_2\frac{M_2-1}{2}\right)\right] &\leq \frac{B_{22}}{\sqrt{n}},\\
K_3\frac{(M_1-1)(M_2-1)}{2}P_{X_1^n}P_{X_2^n}Q^{(3)}_{Y^n}\!\left[\,\tilde{i}(X_1^nX_2^n;Y^n)\!\!>\!\log\left(K_3\frac{(M_1-1)(M_2-1)}{2}\right)\right] &\leq \frac{B_{32}}{\sqrt{n}}.
\end{align}
\label{conf-MAC}
\end{subequations}

\subsubsection{Completion}

Substituting~\eqref{cost=0-MAC},~\eqref{outage-mac-joint}, and~\eqref{conf-MAC} into the achievability bound~\eqref{DT-joint} of Theorem~\ref{corol-modified-DT-MAC} and recalling~\eqref{err-mac} that, with a little abuse of notation cf.~\cite[Eq. (186)]{PPV}, $\epsilon$ is the target error probability yields 
\begin{align}
&\epsilon 
\geq 1- \Pr\left[\mathcal{N}(n\mathbf{C}(P_1,P_2),n\mathbf{V}(P_1,P_2))>\begin{pmatrix}
\log\left(K_1\frac{M_1-1}{2}\right) \\
\log\left(K_2\frac{M_2-1}{2}\right) \\
\log\left(K_3\frac{(M_1-1)(M_2-1)}{2}\right)
\end{pmatrix}\right] + \frac{B}{\sqrt{n}}, 
\end{align}
where~$B=B_1+B_{21}+B_{22}+B_{23}$. Rearranging and using the symmetry property of the Gaussian distribution $\Pr[\mathcal{N}>z]=\Pr[\mathcal{N}<-z]$, we obtain
\begin{align}
\Pr\left[\mathcal{N}(\mathbf{0},n\mathbf{V}(P_1,P_2))<n\mathbf{C}(P_1,P_2)-
\begin{pmatrix}
\log\left(K_1\frac{M_1-1}{2}\right) \\
\log\left(K_2\frac{M_2-1}{2}\right) \\
\log\left(K_3\frac{(M_1-1)(M_2-1)}{2}\right)
\end{pmatrix}\right] \geq 1- \left(\epsilon - \frac{B}{\sqrt{n}}\right). 
\end{align}
Recalling the definition~\eqref{Q-inv} of the inverse complementary CDF of the multi-dimensional Gaussian RV, we find
\begin{align}
\begin{pmatrix}
\log\left(\frac{M_1-1}{2}\right) \\
\log\left(\frac{M_2-1}{2}\right) \\
\log\left(\frac{(M_1-1)(M_2-1)}{2}\right) \\
\end{pmatrix}  &\in n\mathbf{C}(P_1,P_2)-\sqrt{n}Q^{-1}\left(\epsilon- \frac{B}{\sqrt{n}}; \mathbf{V}(P_1,P_2)\right) -
\begin{pmatrix}
\log K_1 \\
\log K_2 \\
\log K_3 \\
\end{pmatrix}  \\
&\subseteq n\mathbf{C}(P_1,P_2)-\sqrt{n} Q^{-1}\left(\epsilon; \mathbf{V}(P_1,P_2)\right)  + O(1)\mathbf{1},  \label{Taylor-MAC-1}
\end{align}
where~\eqref{Taylor-MAC-1} follows from the Taylor expansion for the multi-dimensional $Q^{-1}$ function.
Thus, we have proved that an $(n,M_1,M_2,\epsilon,P_1,P_2)$ code exists if the rate pair satisfies
\begin{align}
\frac{1}{n}
\begin{pmatrix}
\log M_1 \\
\log M_2 \\
\log\left(M_1M_2\right) \\
\end{pmatrix}  &\in \mathbf{C}(P_1,P_2)-\frac{1}{\sqrt{n}}Q^{-1}\left(\epsilon; \mathbf{V}(P_1,P_2)\right)+ O\left(\frac{1}{n}\right)\mathbf{1}.
\label{MAC-ach-approx}
\end{align}
This concludes the proof of achievability for the joint-outage rate region of Theorem~\ref{GMAC-joint}.

Next, we turn to the proof of Theorem 3. Substituting~\eqref{cost=0-MAC},~\eqref{outage-mac-splitting}, and~\eqref{conf-MAC} into the achievability bound~\eqref{DT-splitting} of Theorem~\ref{corol-modified-DT-MAC} and again recalling~\eqref{err-mac} that $\epsilon$ is the target error probability leads to 
\begin{align}
\epsilon
&\geq Q\left(\frac{nC(P_1)-\log\left(K_1\frac{M_1-1}{2}\right)}{\sqrt{nV(P_1)}}\right) + \frac{\tilde{B}_1}{\sqrt{n}} \notag\\
&+ Q\left(\frac{nC(P_2)-\log\left(K_2\frac{M_2-1}{2}\right)}{\sqrt{nV(P_2)}}\right) + \frac{\tilde{B}_2}{\sqrt{n}} \notag\\
&+ Q\left(\frac{nC(P_1+P_2)-\log\left(K_3\frac{(M_1-1)(M_2-1)}{2}\right)}{\sqrt{n[V(P_1+P_2)+V_3(P_1,P_2)]}}\right) + \frac{\tilde{B}_3}{\sqrt{n}}, 
\end{align}
where~$\tilde{B}_1:=B_{11}+B_{21}$, $\tilde{B}_2:=B_{12}+B_{22}$, and $\tilde{B}_3:=B_{13}+B_{23}$. Now, splitting $\epsilon$ among the three first terms of each line gives
\begin{align}
&\log M_1\leq nC(P_1)-\sqrt{nV(P_1)}Q^{-1}\left(\lambda_1\epsilon-\frac{\tilde{B}_1}{\sqrt{n}}\right)-\log K_1\notag\\
&\log M_2\leq nC(P_2)-\sqrt{nV(P_2)}Q^{-1}\left(\lambda_2\epsilon-\frac{\tilde{B}_2}{\sqrt{n}}\right)-\log K_2\notag\\
&\log M_1+\log M_2\leq nC(P_1+P_2)-\sqrt{n[V(P_1+P_2)+V_3(P_1,P_2)]}Q^{-1}\left(\!\lambda_3\epsilon-\!\frac{\tilde{B}_3}{\sqrt{n}}\!\right)-\log K_3
\label{bounds on CLT}
\end{align}
where the positive constants $\lambda_1,\lambda_2,\lambda_3$ such that $\lambda_1+\lambda_2+\lambda_3=1$ can be arbitrarily chosen to represent the weight of each of the three types of outage. We can further simplify the bounds in \eqref{bounds on CLT} using the Taylor expansion $Q^{-1}(\lambda\epsilon-\frac{\tilde{B}}{\sqrt{n}})= Q^{-1}(\lambda\epsilon)+O(1/\sqrt{n})$ to obtain
\begin{align}
&\log M_1\leq nC(P_1)-\sqrt{nV(P_1)}Q^{-1}\left(\lambda_1\epsilon\right)+O(1),\notag\\
&\log M_2\leq nC(P_2)-\sqrt{nV(P_2)}Q^{-1}\left(\lambda_2\epsilon\right)+O(1),\notag\\
&\log M_1+\log M_2\leq nC(P_1+P_2)-\sqrt{n[V(P_1+P_2)+V_3(P_1,P_2)]}Q^{-1}\left(\lambda_3\epsilon\right)+O(1).
\end{align}
This concludes the proof of achievability for the outage-splitting  rate region of Theorem~\ref{GMAC-splitting}.

\newpage

\section{Conclusion}
\label{sec-con}

We have proved several inner bounds for the Gaussian MAC in the finite blocklength regime, and used them to establish second-order achievable rate regions. As a consequence of our study, we observe that codebooks that are randomly generated according to independent uniform distributions on the users' power shells result in rather tight second-order rate regions for the Gaussian MAC, and they outperform coding schemes induced by the (first-order-optimal) Gaussian input distribution and those via TDMA.

To obtain these main results, we have developed simple and transparent methods for proving non-asymptotic achievability results for Gaussian settings. Our achievability methods rely on the conventional random coding and typicality decoding, but employs modified mutual information random variables, a new change of measure technique, and the application of a CLT for functions. We believe that our methods provide valuable insights for handling other channel models involving input cost constraints, and they may also be generalized to other multi-user settings.



%

\appendices

\section{Proof of Proposition~\ref{CLT-func}: CLT for Functions}
\label{proof-Prop-P2P}

Since the vector-valued function~$\mathbf{f}(\mathbf{u})$ has continuous second-order partial derivatives at $\mathbf{0}$, we have from Taylor's Theorem that
\begin{align}
\mathbf{f}\left(\mathbf{u}\right)=\mathbf{f}(\mathbf{0}) + \mathbf{J} \mathbf{u}^T + \mathbf{R}(\mathbf{u}),
\end{align}
where $\mathbf{R}(\mathbf{u})$ is the vanishing remainder term in the Taylor expansion. In particular, for those $\mathbf{u}$ belonging to the $K$-hypercube neighborhood~$N(r_0)$ of $\mathbf{0}$ with side length~$r_0>\frac{1}{\sqrt[4]{n}}$, the Lagrange (mean-value) form of the Taylor Theorem provides the following uniform bound on the remainder term 
\begin{align}
|\mathbf{R}(\mathbf{u})| \leq \frac{1}{2}
\begin{bmatrix}
\max_{1\leq k,p\leq K} \max_{\mathbf{u}_0\in N(r_0)} \left|\frac{\partial^2 f_1(\mathbf{u}_0)}{\partial u_k \partial u_p}\right| \\
\cdot \\
\cdot \\
\cdot \\
\max_{1\leq k,p\leq K} \max_{\mathbf{u}_0\in N(r_0)} \left|\frac{\partial^2 f_L(\mathbf{u}_0)}{\partial u_k \partial u_p}\right|
\end{bmatrix}
(u_1+...+u_K)^2, \label{lag-bound}
\end{align}
where $|\mathbf{u}|:=(|u_1|,...,|u_K|)$ denotes the element-wise absolute value, and the vector inequality in~\eqref{lag-bound} is also element wise.

Now, we apply the normalized sum $\frac{1}{n}\sum_{t=1}^n \mathbf{U}_t$ as the argument of function~$\mathbf{f}$ to obtain
\begin{align}
\mathbf{f}\left(\frac{1}{n}\sum_{t=1}^n \mathbf{U}_t\right)=\mathbf{f}(\mathbf{0}) + \mathbf{J} \frac{1}{n}\sum_{t=1}^n \mathbf{U}_t^T + \mathbf{R}\left(\frac{1}{n}\sum_{t=1}^n \mathbf{U}_t\right) \label{taylor}
\end{align}
almost surely. Since the random vector~$\frac{1}{n}\sum_{t=1}^n \mathbf{U}_t$ is concentrating around $\mathbf{0}$, we conclude that the corresponding remainder term also concentrates around~$0$:
\begin{align}
&\Pr\left[\left|\mathbf{R}\left(\frac{1}{n}\sum_{t=1}^n \mathbf{U}_t\right)\right|>\frac{1}{\sqrt{n}}\mathbf{1}\right]  \notag \\
&\;  \leq \Pr\left[\left|\mathbf{R}\left(\frac{1}{n}\sum_{t=1}^n \mathbf{U}_t\right)\right|>\frac{1}{\sqrt{n}}\mathbf{1} \bigcap \frac{1}{n}\sum_{t=1}^n \mathbf{U}_t \in N(r_0)\right] 
+ \Pr\left[\frac{1}{n}\sum_{t=1}^n \mathbf{U}_t \notin N(r_0)\right] \label{neigh} \\
&\; \leq \Pr\left[\frac{1}{2}
\begin{bmatrix}
\max_{1\leq k,p\leq K} \max_{\mathbf{u}_0\in B(r_0)} \left|\frac{\partial^2 f_1(\mathbf{u}_0)}{\partial u_k \partial u_p}\right| \\
\cdot \\
\cdot \\
\cdot \\
\max_{1\leq k,p\leq K} \max_{\mathbf{u}_0\in B(r_0)} \left|\frac{\partial^2 f_L(\mathbf{u}_0)}{\partial u_k \partial u_p}\right|
\end{bmatrix}
\left(\frac{1}{n}\sum_{t=1}^n (U_{1t}+...+U_{Kt})\right)^2>\frac{1}{\sqrt{n}}\mathbf{1}\right] 
+\sum_{k=1}^K \Pr\left[\left|\frac{1}{n}\sum_{t=1}^n U_{kt}\right| >r_0\right] \label{Lagrange} \\
&\; = \Pr\left[ \left(\frac{1}{n}\sum_{t=1}^n (U_{1t}+...+U_{Kt})\right)^2>\frac{2}{\sqrt{n}} \left(\min_{1\leq l\leq L} \max_{1\leq k,p\leq K} \max_{\mathbf{u}_0\in B(r_0)} \left|\frac{\partial^2 f_l(\mathbf{u}_0)}{\partial u_k \partial u_p}\right|\right)^{-1}\right] +\sum_{k=1}^K \Pr\left[\left|\frac{1}{n}\sum_{t=1}^n U_{kt}\right| >r_0\right] \\
&\; \leq \frac{\text{Var}\left[\frac{1}{n}\sum_{t=1}^n (U_{1t}+...+U_{Kt})\right]}{\frac{2}{\sqrt{n}} \left(\min_{1\leq l\leq L} \max_{1\leq k,p\leq K} \max_{\mathbf{u}_0\in B(r_0)} \left|\frac{\partial^2 f_l(\mathbf{u}_0)}{\partial u_k \partial u_p}\right|\right)^{-1}} +\sum_{k=1}^K \frac{\text{Var}\left[\frac{1}{n}\sum_{t=1}^n U_{kt}\right]}{r_0^2} \label{cheb} \\
&\; \leq \frac{\frac{K}{n}\left(\text{Var}[U_{11}]+...+\text{Var}[U_{K1}]\right)}{\frac{2}{\sqrt{n}} \left(\min_{1\leq l\leq L} \max_{1\leq k,p\leq K} \max_{\mathbf{u}_0\in B(r_0)} \left|\frac{\partial^2 f_l(\mathbf{u}_0)}{\partial u_k \partial u_p}\right|\right)^{-1}} + \frac{\text{Var}[U_{11}]+...+\text{Var}[U_{K1}]}{nr_0^2} \label{sum-var} \\
&\; = \frac{c_1}{\sqrt{n}}, \label{c1}
\end{align}
where~\eqref{neigh} follows from the simple bound~$\Pr[A]\leq \Pr[A\cap B]+\Pr[B^c]$,~\eqref{Lagrange} from the Lagrange bound~\eqref{lag-bound} and the union bound,~\eqref{cheb} follows from the Chebyshev inequality,~\eqref{sum-var} from the simple bound on the sum of variances of generic dependent random variables~$\text{Var}[X_1+...+X_K]\leq K(\text{Var}[X_1]+...+\text{Var}[X_K])$, and~\eqref{c1} from the constraint~$r_0>\frac{1}{\sqrt[4]{n}}$ on the side length of the neighborhood and the definition of the constant $c_1$ as 
\begin{align}
c_1:=\left(\text{Var}[U_{11}]+...+\text{Var}[U_{K1}]\right) \left[1+\frac{K}{2}\min_{1\leq l\leq L} \max_{1\leq k,p\leq K} \max_{\mathbf{u}_0\in B(r_0)} \left|\frac{\partial^2 f_l(\mathbf{u}_0)}{\partial u_k \partial u_p}\right|\right]. \label{c1-def}
\end{align}
In the derivation above, we have assumed that~$\min_{1\leq l\leq L} \max_{1\leq k,p\leq K} \max_{\mathbf{u}_0\in B(r_0)} \left|\frac{\partial^2 f_l(\mathbf{u}_0)}{\partial u_k \partial u_p}\right|>0$; in case this does not hold, that is, if~$\min_{1\leq l\leq L} \max_{1\leq k,p\leq L} \max_{\mathbf{u}_0\in B(r_0)} \left|\frac{\partial^2 f_l(\mathbf{u}_0)}{\partial u_k \partial u_p}\right|=0$, the above sequence of steps do not hold, but the final result~\eqref{c1} trivially holds with $c_1=\text{Var}[U_{11}]+...+\text{Var}[U_{K1}]$ which is consistent with~\eqref{c1-def}.

Now, we obtain
\begin{align}
&\Pr\left[\mathbf{f}\left(\frac{1}{n}\sum_{t=1}^n \mathbf{U}_t\right) \in \mathcal{D}\right] \notag \\
&\qquad  \leq \Pr\left[\mathbf{f}\left(\frac{1}{n}\sum_{t=1}^n \mathbf{U}_t\right) \in \mathcal{D} \; \bigcap \; \left|\mathbf{R}\left(\frac{1}{n}\sum_{t=1}^n \mathbf{U}_t\right)\right|\leq\frac{1}{\sqrt{n}}\mathbf{1} \right] + \Pr\left[\left|\mathbf{R}\left(\frac{1}{n}\sum_{t=1}^n \mathbf{U}_t\right)\right|>\frac{1}{\sqrt{n}}\mathbf{1} \right] \label{step-0} \\
&\qquad  \leq \Pr\left[ \mathbf{f}(\mathbf{0})+\frac{1}{n}\sum_{t=1}^n \mathbf{J}\mathbf{U}_t^T \in \mathcal{D}\oplus\frac{1}{\sqrt{n}}\mathbf{1} \right] + \frac{c_1}{\sqrt{n}} \label{step-1} \\
&\qquad  \leq  \Pr\left[ \mathcal{N}\left( \mathbf{f}(\mathbf{0})+\mathbb{E}[\mathbf{J}\mathbf{U}_1^T] , \frac{1}{n}\text{Cov}[\mathbf{J}\mathbf{U}_1^T]\right) \in \mathcal{D}\oplus\frac{1}{\sqrt{n}}\mathbf{1} \right] + \frac{c_2}{\sqrt{n}} + \frac{c_1}{\sqrt{n}} \label{step-2} \\
&\qquad  \leq  \Pr\left[ \mathcal{N}\left(\mathbf{f}(\mathbf{0}),\frac{1}{n}\mathbf{J}\text{Cov}[\mathbf{U}_1]\mathbf{J}^T\right) \in \mathcal{D} \right] + \frac{c_3}{\sqrt{n}} + \frac{c_2}{\sqrt{n}} + \frac{c_1}{\sqrt{n}} \label{step-3}
\end{align}
where inequality~\eqref{step-0} follows from the simple bound~$\Pr[A]\leq \Pr[A\cap B]+\Pr[B^c]$,~\eqref{step-1} from~\eqref{taylor} and~\eqref{c1} as well as the definition of a ``linear outward set-expansion''~$\mathcal{D}\oplus\frac{1}{\sqrt{n}}\mathbf{1}$, which is closely related to the formal definition of set expansion in~\cite{set-expansion} and basically means an enlargement of the set~$\mathcal{D}$ with an ``addition in all directions'' with~$\frac{1}{\sqrt{n}}$,~\eqref{step-2} from the multi-dimensional CLT~\cite{3D-CLT,Tan} with the constant~$c_2$ defined as
\begin{align}
\tilde{c}_2:=\frac{400\,L^{1/4} \mathbb{E}[||\mathbf{J}\mathbf{U}_1^T||_2^3]}{\lambda_{\text{min}}\left(\text{Cov}[\mathbf{J}\mathbf{U}_1^T]\right)^{3/2}}\leq \frac{400\,L^{1/4} \lambda_{\text{max}}\left(\mathbf{J}\mathbf{J}^T\right)^{3/2}\mathbb{E}[||\mathbf{U}_1^T||_2^3]}{\lambda_{\text{min}}\left(\text{Cov}[\mathbf{J}\mathbf{U}_1^T]\right)^{3/2}}:=c_2,
\end{align}
where $\lambda_{\text{min}}(\Sigma)$ and $\lambda_{\text{max}}(\Sigma)$ denotes the smallest and largest eigenvalues of the matrix~$\Sigma$, respectively, and finally \eqref{step-3} from the Taylor expansion for the probability at hand with the proper positive finite constant~$c_3$ depending upon the set~$\mathcal{D}$. 

Analogously, we have
\begin{align}
\Pr\left[\mathbf{f}\left(\frac{1}{n}\sum_{t=1}^n \mathbf{U}_t\right) \in \mathcal{D}\right]
& \geq \Pr\left[\mathbf{f}\left(\frac{1}{n}\sum_{t=1}^n \mathbf{U}_t\right) \in \mathcal{D} \; \bigcap \; \left|\mathbf{R}\left(\frac{1}{n}\sum_{t=1}^n \mathbf{U}_t\right)\right|\leq\frac{1}{\sqrt{n}}\mathbf{1} \right]  \\
&\geq \Pr\left[\mathbf{f}(\mathbf{0})+\mathbf{J} \frac{1}{n}\sum_{t=1}^n \mathbf{U}_t^T  \in \mathcal{D}\ominus\frac{1}{\sqrt{n}}\mathbf{1} \; \bigcap \; \left|\mathbf{R}\left(\frac{1}{n}\sum_{t=1}^n \mathbf{U}_t\right)\right|\leq\frac{1}{\sqrt{n}}\mathbf{1} \right] \label{set-cont} \\
& \geq \Pr\left[\mathbf{f}(\mathbf{0})+ \frac{1}{n}\sum_{t=1}^n \mathbf{J}\mathbf{U}_t^T \in \mathcal{D}\ominus\frac{1}{\sqrt{n}}\mathbf{1} \right] - \Pr\left[\left|\mathbf{R}\left(\frac{1}{n}\sum_{t=1}^n \mathbf{U}_t\right)\right|>\frac{1}{\sqrt{n}}\mathbf{1} \right] \label{step-11} \\
& \geq \Pr\left[ \mathbf{f}(\mathbf{0})+\frac{1}{n}\sum_{t=1}^n \mathbf{J}\mathbf{U}_t^T \in \mathcal{D}\ominus\frac{1}{\sqrt{n}}\mathbf{1} \right] - \frac{c_1}{\sqrt{n}} \\
& \geq  \Pr\left[ \mathcal{N}\left( \mathbf{f}(\mathbf{0})+\mathbb{E}[\mathbf{J}\mathbf{U}_1^T] , \frac{1}{n}\text{Cov}[\mathbf{J}\mathbf{U}_1^T]\right) \in \mathcal{D}\ominus\frac{1}{\sqrt{n}}\mathbf{1} \right] - \frac{c_2}{\sqrt{n}} - \frac{c_1}{\sqrt{n}} \\
& \geq  \Pr\left[ \mathcal{N}\left(\mathbf{f}(\mathbf{0}),\frac{1}{n}\mathbf{J}\text{Cov}[\mathbf{U}_1]\mathbf{J}^T\right) \in \mathcal{D} \right] - \frac{c_3}{\sqrt{n}} - \frac{c_2}{\sqrt{n}} - \frac{c_1}{\sqrt{n}} \label{step-13}
\end{align}
where inequality~\eqref{set-cont} follows from the definition of a ``linear inward set-contraction''~$\mathcal{D}\ominus\frac{1}{\sqrt{n}}\mathbf{1}$, which is closely related to the formal definition of set contraction in~\cite{set-expansion} and basically means a shrinkage of the set~$\mathcal{D}$ with a ``deduction in all directions'' of~$\frac{1}{\sqrt{n}}$,~\eqref{step-11} follows from the bound~$\Pr[A\cap B]\geq \Pr[A]-\Pr[B^c]$, and all the other steps are as in the previous case. 

Combining inequalities~\eqref{step-3} and~\eqref{step-13} establishes Proposition~\ref{CLT-func} with the constant $B:=c_1+c_2+c_3$.

\section{Proof of Proposition~\ref{Prop-P2P-unif-bound} for P2P Gaussian Channels}
\label{proof-Prop-P2P}

Define $D_{P,Q}(y^n):=\frac{P_{Y^n}(y^n)}{Q_{Y^n}(y^n)}$. Recalling the output distribution~\eqref{shell-output} induced by the uniform input distribution on the power shell~\eqref{unif-shell}, we can simplify $D_{P,Q}(y^n)$ as
\begin{align}
D_{P,Q}(y^n)&=\frac{1}{2}\left(2e^{-P}(1+P)\right)^{n/2}\Gamma\left(\frac{n}{2}\right)e^{-P||y^n||^2/2(1+P)}  \frac{I_{n/2-1}(||y^n||\sqrt{nP})}{(||y^n||\sqrt{nP})^{n/2-1}}.
\end{align}
To bound this divergence, we first notice that 
\begin{align}
\ln \Gamma\left(\frac{n}{2}\right)\leq\frac{n-1}{2}\ln\left(\frac{n}{2}\right)-\frac{n}{2}+c_\Gamma. \label{Gamma}
\end{align}
where $c_\Gamma\leq 2$; in fact, for asymptotically large $n$, the above inequality tends to equality with $c_\Gamma=\ln(\sqrt{2\pi})$ due to Sterling's approximation. 
Moreover, $I_{k}(z)\leq I_{k+1}(z)$ for any order~$k$, and so it is sufficient to bound the above divergence only for even values of the order, such that $k=n/2-1$ is an integer. For such an integer, we have~\cite{PPV}
 \begin{align}
z^{-k}I_{k}(z)\leq \sqrt{\frac{\pi}{8}}\left(k^2+z^2\right)^{-1/4}\left(k+\sqrt{k^2+z^2}\right)^{-k}e^{\sqrt{k^2+z^2}} \label{bessel-bound}
\end{align}
Using the above inequality along with shorthands $a=\frac{n/2-1}{n/2}$, we obtain
\begin{align}
\ln D_{P,Q}(y^n)&\leq c+\frac{n}{2}f_{a,P}\left(\frac{||y^n||^2}{n}\right),
\end{align}
where $c=\ln(1/2)+c_\Gamma+\ln(\sqrt{\pi/8})=O(1)$, and for $t\in\mathbb{R}^+$
\begin{align}
f_{a,P}(t)&:=\ln\left(2e^{-(1+P)}(1+P)\right)-\frac{Pt}{1+P}+\sqrt{a^2+4Pt} -a\ln\left(a+\sqrt{a^2+4Pt}\right)-\frac{1-a}{2}\ln\left(\sqrt{a^2+4Pt}\right).
\end{align}
To prove the proposition for any finite~$n$, one needs to show that the above function~$f_{a,P}(t)$ is non-positive for all~$t\in\mathbb{R}^+$, for any fixed~$P$. For simplicity, however, we only focus on sufficiently large values of~$n$, such that $a\to1$. In such a case, the above function simplifies to 
\begin{align}
f_P(t)&:=\ln\left(2e^{-(1+P)}(1+P)\right)-\frac{Pt}{1+P} +\sqrt{1+4Pt}-\ln\left(1+\sqrt{1+4Pt}\right).
\end{align}
It is easy to show that the function~$f_P(t)$ has only one local (and also global) maximum which occurs at $t=1+P$ leading to~$f_P(1+P)=0$. Therefore,~$f_P(t)\leq0$ for all~$t\in\mathbb{R}^+$, concluding that for all $y^n\in\mathbb{R}^n$
\begin{align}
D_{P,Q}(y^n)
&\leq \text{exp}\left(c+\frac{n}{2}f_P\left(\frac{||y^n||^2}{n}\right)\right) \leq K,
\end{align}
where $K:=e^c\leq 1$. Notice that, interestingly, the constant~$K$ on the RHS of this inequality is independent of the power constraint~$P$.

\section{Proof of Proposition~\ref{divergence-bound-MAC} for the Gaussian MAC}
\label{proof-Prop-MAC}

Proof is similar to that of Proposition 2. The first two inequalities~\eqref{unif-mac-1} and~\eqref{unif-mac-2} indeed directly follow from Proposition~1, since the conditional outputs~$P_{Y^n|X_2^n}(y^n|x_2^n),P_{Y^n|X_1^n}(y^n|x_1^n)$ induced by the power shell distribution and the per-user capacity achieving distributions~$Q^{(1)}_{Y^n|X_2^n}(y^n|x_2^n),Q^{(2}_{Y^n|X_1^n}(y^n|x_1^n)$ both have the same expressions as the output distribution~\eqref{shell-output} induced by the P2P shell distribution and the capacity-achieving output distribution of a P2P channel, respectively, with the only modification that $y^n\in\mathbb{R}^n$ is replace by $(y^n-x_2^n)\in\mathbb{R}^n$ and $(y^n-x_1^n)\in\mathbb{R}^n$, and $P$ by $P_1$ and $P_2$, respectively. 

Therefore, it only remain to prove the third inequality~\eqref{unif-mac-3} on the unconditional R-N derivative $\frac{P_{Y^n}(y^n)}{Q^{(3)}_{Y^n}(y^n)}$. Since the output distribution~$P_{Y^n}$ is not explicitly available, we take an indirect approach. We show that the corresponding \emph{input} distributions satisfy the desired property and thus conclude that their resulting output distribution do as well. In particular, let $Q_{U^n}(u^n)\sim\mathcal{N}(u^n; \mathbf{0},(P_1+P_2)I_n)$ be the distribution of the superimposed input~$U^n=X_1^n+X_2^n$ when the two inputs~$X_1^n$ and~$X_2^n$ are independent i.i.d. Gaussian distributions. Note that feeding this distribution to the channel $Y^n=U^n+Z^n$ recovers the capacity-achieving output distribution~$Q^{(3)}_{Y^n}(y^n)\sim\mathcal{N}(y^n; \mathbf{0},(1+P_1+P_2)I_n)$:
\begin{align}
Q^{(3)}_{Y^n}(y^n) =\int_{\mathbb{R}^n}  Q_{U^n}(u^n) P_{Y^n|U^n}(y^n|u^n)  du^n. 
\end{align}
Therefore, if we can show that 
\begin{align}
D_{P,Q}(u^n):=\frac{P_{U^n}(u^n)}{Q_{U^n}(u^n)} \leq K_3, \qquad \forall u^n\in\mathbb{R}^n \label{unif-mac-input}
\end{align}
then it immediately follows for any $y^n\in\mathbb{R}^n$ that
\begin{align}
P_{Y^n}(y^n) &= \int_{\mathbb{R}^n}  P_{U^n}(u^n) P_{Y^n|U^n}(y^n|u^n)  du^n \\
&\leq \int_{\mathbb{R}^n}  K_3\, Q_{U^n}(u^n) P_{Y^n|U^n}(y^n|u^n)  du^n= K_3\, Q^{(3)}_{Y^n}(y^n) . 
\end{align}
Hence, we are only left with the proof of~\eqref{unif-mac-input}. Note that, the claim is trivial for those $u^n\in\mathbb{R}^n$ not belonging to the hollow sphere $|\sqrt{nP_1}-\sqrt{nP_2}|< ||u^n|| < \sqrt{nP_1}+\sqrt{nP_2}$, since they satisfy $P_{U^n}(u^n)=0$. Thus, focusing on those~$u^n$ belonging to this hollow sphere, we have
\begin{align}
D_{P,Q}(u^n)
&=  \sqrt{\frac{P_2}{\pi P_1}} \frac{\Gamma\left(\frac{n}{2}\right)}{||u^n||\Gamma\left(\frac{n-1}{2}\right)} \frac{\Gamma\left(\frac{n}{2}\right)(2\pi)^{n/2}(P_1+P_2)^{n/2} e^{||u^n||^2/2(P_1+P_2)}}{2\pi^{n/2}(nP_2)^{(n-1)/2}} \left(1-\left(\frac{||u^n||^2+n(P_1-P_2)}{2\sqrt{nP_1}||u^n||}\right)^2\right)^{(n-3)/2}
\end{align}
Using~\eqref{Gamma} and the crude bound $\Gamma\left(\frac{n}{2}\right)/\Gamma\left(\frac{n-1}{2}\right)\leq \sqrt{n}$, we obtain
\begin{align}
\ln D_{P,Q}(u^n) &\leq  \ln\left(\frac{P_2}{\sqrt{\pi P_1}}\right) + \ln \left(\frac{n}{2}\right) - \ln(||u^n||) + \frac{n-1}{2}\ln\left(\frac{n}{2}\right)-\frac{n}{2}+c_\Gamma + \frac{n}{2}\ln\left(\frac{2(P_1+P_2)}{P_2}\right) - \frac{n}{2}\ln(n) \notag \\
& \qquad \qquad + \frac{||u^n||^2}{2(P_1+P_2)} + \frac{n-3}{2}\ln \left(1-\left(\frac{||u^n||^2+n(P_1-P_2)}{2\sqrt{nP_1}||u^n||}\right)^2\right) \\
&=  c + \frac{n}{2} f_{n,P_1,P_2}\left(\frac{||u^n||^2}{n}\right) 
\end{align}
where $c:=c_\Gamma + \ln\left(\frac{P_2}{\sqrt{2\pi P_1}}\right)$ and 
\begin{align}
f_{n,P_1,P_2}(t)=& - \frac{\ln(t)}{n} + \ln\left(\frac{P_1+P_2}{e\,P_2}\right) + \frac{t}{P_1+P_2} + \frac{n-3}{n}\ln \left(1-\frac{(t+P_1-P_2)^2}{4P_1t}\right),
\end{align}
with $(\sqrt{P_1}-\sqrt{P_2})^2 < t < (\sqrt{P_1}+\sqrt{P_2})^2$. 
To prove the proposition for any finite~$n$, one needs to show that the above function~$f_{n,P_1,P_2}(t)$ is non-positive for all~$t$ in the aforementioned range, for any fixed~$P_1,P_2$. For simplicity, however, we only focus on sufficiently large values of~$n$. In such a case, the above function simplifies to 
\begin{align}
f_{P_1,P_2}(t)=& \ln\left(\frac{P_1+P_2}{e\,P_2}\right) + \frac{t}{P_1+P_2} +\ln \left(1-\frac{(t+P_1-P_2)^2}{4P_1t}\right).
\end{align}
It is then easy to show that, in this range of values for $t$, the function~$f_{P_1,P_2}(t)$ has only one local (and also global) maximum which occurs at $t=P_1+P_2$ leading to~$f_{P_1,P_2}(P_1+P_2)=0$. Therefore,~$f_{P_1,P_2}(t)\leq0$ for all~$t\in((\sqrt{P_1}-\sqrt{P_2})^2 , (\sqrt{P_1}+\sqrt{P_2})^2)$, concluding that, for any $u^n$ satisfying $|\sqrt{nP_1}-\sqrt{nP_2}|< ||u^n|| < \sqrt{nP_1}+\sqrt{nP_2}$,
\begin{align}
D_{P,Q}(u^n) &\leq \text{exp}\left(c+\frac{n}{2}f_{P_1,P_2}\left(\frac{||u^n||^2}{n}\right)\right) \leq K_3,
\end{align}
where $K_3:=e^c = \frac{e^{c_\Gamma}\,P_2}{\sqrt{2\pi P_1}}=O(1)$. This concludes the proof of Proposition 3. Note that, in the case of Gaussian MAC, the constant~$K_3$  \textit{depends} upon the power constraints~$P_1$ and~$P_2$, at least as indicated by our bounding techniques.

\section*{Acknowledgment}

This work has beeb supported in part by the NSF grants CCF05-46618 and CPS12-39222.

\ifCLASSOPTIONcaptionsoff
  \newpage
\fi

\end{document}